\documentclass[aps,prd,twocolumn,preprintnumbers,superscriptaddress,nofootinbib,floatfix]{revtex4-1}
\usepackage{mathrsfs}
\usepackage{amsmath}
\usepackage{slashed}
\usepackage{graphicx,color}
\usepackage{dcolumn}
\usepackage{bm}
\usepackage{subfigure}
\usepackage{graphicx}
\usepackage{amssymb}
\usepackage{stackrel}
\usepackage{pgfplots}
\usetikzlibrary{positioning}
\usetikzlibrary{snakes}
\usepackage{hyperref}
\usepackage{comment}
\usepackage{amsfonts}
\usepackage{epstopdf}
\usepackage{hyperref}
\usepackage{array}
\usepackage[utf8]{inputenc}
\usepackage{soul}
\usepackage{color}
\usepackage[T1]{fontenc}
\usepackage{float}

\newcommand{\be}{\begin{equation}}
\newcommand{\ee}{\end{equation}}
\newcommand{\bea}{\begin{eqnarray}}
\newcommand{\eea}{\end{eqnarray}}

\begin{document}

\title{Direct detection of Sub-GeV Dark Matter via 3-body Inelastic Scattering Process}

\author{Wei Chao}
\email{chaowei@bnu.edu.cn}
\author{Mingjie Jin}
\email{jinmj@bnu.edu.cn}
\author{Ying-Quan Peng}
\email{yqpenghep@mail.bnu.edu.cn}
\affiliation{Center for Advanced Quantum Studies, Department of Physics, Beijing Normal University, Beijing, 100875, China}
\vspace{3cm}

\begin{abstract}

Direct detection of Sub-GeV dark matter (DM) is challenging because the recoil energy of the nuclei or electron from the elastic scattering of a sub-GeV DM off the target can hardly reach the detector threshold.  In this paper, we present a new direct detection strategy for sub-GeV DM  via the three-body inelastic scattering process,  $\chi + \chi + {\rm SM} \to \eta + {\rm SM}$, where $\chi$ is DM candidate and $\eta $ is either a DM composite state or any dark radiation.  This process is common for a large class of DM models without presuming  particular thermal history in the early Universe. The typical signature from this process is almost a monoenergetic pulse signal where the recoil energy  comes from either the binding energy or the  consumed DM particle. We show that detectable DM mass range  can be effectively enlarged compared to the elastic scattering process.

\end{abstract}

\maketitle


\section{Introduction}

Despite ample evidence of dark matter (DM) in our Universe,  its particle nature (mass, spin  and coupling) is still a mystery.  The arbitrary mass of possible DM, ranging form $10^{-22}$~eV to $10^{55}$ GeV,  leaves the experimental observation of DM a severe challenge. Many ideas have been proposed to search for DM in laboratory~\cite{Liu:2017drf,Lin:2019uvt}.  Assuming DM origins from thermal processes in the primordial plasma, its mass range is limited to $ [{\cal O}(1)~{\rm keV},~{\cal O}(100)~{\rm TeV}]$, where the upper bound is required by the  unitarity constraint on DM annihilation amplitude~\cite{Griest:1989wd} and the lower bound is required by the large scale structure of the Universe~\cite{Steigman:1984ac}. This kind of DM can be directly detected in underground laboratory by looking for kinetic energy deposited by DM scattering on atomic nuclei.  The direct detections for DM have reached great sensitivities, and give the current most stringent limits for DM masses above a few GeV~\cite{XENON:2018bec,PandaX-II:2017jmq}. However, the traditional direct detection of DM via $2\to2$ elastic scattering process loses sensitivity rapidly for a sub-GeV DM because  the recoil energy  turns to be smaller than detector threshold.

Many new approaches have been proposed for the direct detection of sub-GeV DM,  for example, boosted DM via various cosmic rays~\cite{Chao:2021vja,Das:2021lcr,Jho:2021rmn,Su:2020zny,Fornal:2020npv}; inelastic DM scattering off target nuclei, during which additional excitations are created~\cite{An:2020tcg,He:2020wjs,Baryakhtar:2020rwy, Song:2021yar,ATLAS:2021kog,He:2020sat,Harigaya:2020ckz,Jacobsen:2021vbr,Borah:2020jzi,Chao:2020yro,Dutta:2021wbn,Keung:2020uew,Aboubrahim:2020iwb}; the absorption of a fermion DM by target and emit a nearly massless neutrino~\cite{Dror:2019onn,Dror:2019dib,Dror:2020czw}, which  produces a characteristic signal; searching DM using condensed matter system~\cite{Kahn:2021ttr,Liang:2021zkg,Andersson:2020uwc,Graham:2012su,Essig:2015cda,Lazanu:2012fi}, where excitations in condensed matter systems provides promising signals; large energy transfer induced by  de-excitation of targets~\cite{Lehnert:2019tuw}; considering the Migdal effect~\cite{Essig:2019xkx,Knapen:2020aky,Ibe:2017yqa,Baxter:2019pnz,Bell:2021zkr}, which results in an inelastically excited electron; detecting DM in superconductors~\cite{Hochberg:2021ymx}, which stand out with the lowest threshold, etcetera. So far, these approaches may reach to very low mass  regime, but  some of them are only applicable to specific DM models. New strategy is still needed to  detect  lower mass regime.

In this Letter, we propose a new direct detection strategy for sub-GeV DM  via the three-body inelastic scattering process,  $\chi + \chi + {\rm SM} \to \eta + {\rm SM}$, where $\chi$ is DM candidate and $\eta $ is either a DM composite state or any dark radiation.  If $\eta$ is a bound state of DM, then its binding energy and kinetic energy can be transferred to the target during the scattering, which results in enhanced recoil energy. Alternatively, if $\eta$ is a dark radiation, then two DM masses are consumed during the scattering, which significantly improve the recoil energy.  This scenario is similar to the case of Co-SIMP~\cite{Smirnov:2020zwf}, but  they are essentially different as this scenario does not depend on specific thermal history of DM.  Although the scattering cross section is small, the high DM density as well as the high SM target density make this process possible.

In the following, we first describe the setup and main features of $3\to 2$ process. Then we present two specific DM models and study their 3-body scatterings to explain the excess in electron recoil events given by the XENON1T experiment.  Following that,  the effect induced by the 3-body DM-nuclei scattering is also discussed.  Finally we summarize our main conclusion. Details of our calculation are given in the supplement material.

\section{$3\to 2 $  processes in  DM direct detections}

Previous studies of DM direct detection mainly focus on  $2\to2$ process which is either elastic or inelastic. In this section, we consider the signal of the  $3\to2$   DM-target scattering process. The $3\to 2$ process has been widely applied to address the spectroscopy and relic density of the strongly interacting massive particles (SIMP)~\cite{Kuflik:2015isi,Hochberg:2014dra,Hochberg:2014kqa,Cline:2017tka,Smirnov:2020zwf}, but this process is not dedicatedly studied  in DM direct detections. Compared with the $2\to2$ process, the target nuclei receive more recoil energy from the $3\to 2$ process. Thus this process is more suitable for the direct detection of sub-GeV DM. In general, the recoil energy in $2\to2$ elastic scattering process is $E_R^{\rm max}=2\mu^2 v^2/m_N$ for nucleus and $E_R^{\rm max} \approx \frac{1}{2}m_\chi v^2$ for electron, where $m_\chi, m_N$ denote the mass of DM and nucleus respectively, $\mu$ is the reduced mass of the DM and nucleus, and $v\approx 10^{-3}$ being the initial velocity of DM.  In $3\to2$ inelastic process, the transferred energy to the target particle is,
\be
\Delta E = \frac{(4-\xi^2)m^2_\chi}{2(m_T+2m_\chi)}, \label{recoiler}
\ee
where we have neglected the initial kinetic energy of DM, $\xi \equiv m_f/m_\chi$ denoting the mass ratio of final and initial dark sectors, and $m_T$ is the target mass.  As a result, the recoil energy for electron is $E_R=\Delta E - |E_B|$, where $E_B$ is the binding energy of electron, while $E_R=\Delta E$ for nucleus.

For a real case of the xenon detector,  we consider 11 different binding energy of the electron in xenon shell $(n,l)$, thus the actual recoil energy for electron is determined by $E_{R,i}=E^{\rm max}_R- |E_B|_i$ ($i=1s^2,...5p^6$), where corresponding binding energies are given in the Tab.~\ref{table:bindenergy}.
\begin{figure}[t]
  \centering
  \includegraphics[width=0.45\textwidth]{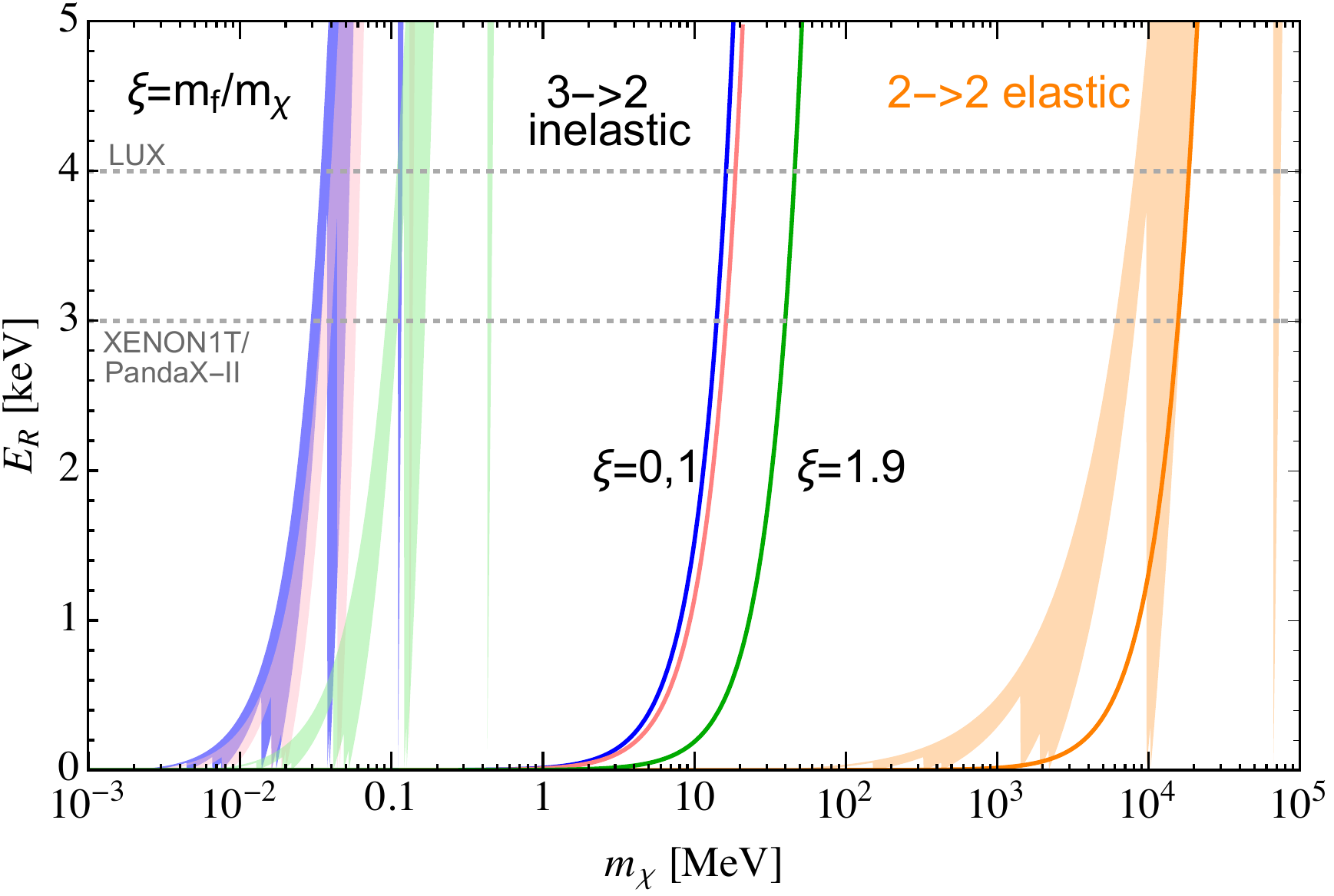}
\caption{The recoil energy $E_R$ vs. $m_\chi$ in $2\to2$ elastic and $3\to2$ inelastic scattering processes. The orange solid line denotes the $2\to2$ process for nucleus, and the blue, pink, green lines describe $3\to2$ process for nucleus with $\xi=0,1,1.9$, respectively. Same color labeled but instead of shaded bands describe $2\to2$ elastic and $3\to2$ inelastic processes for electron. The horizontal gray dashed lines represent the detector threshold of xenon target experiments.}
\label{fig:mchi_er}
\end{figure}
We show in Fig.~\ref{fig:mchi_er} the typical recoil energy $E_R$ as a function of $m_\chi$ in $2\to2$ and $3\to2$  scattering processes, respectively. The orange solid line denotes recoil energy of nucleus from  $2\to2$ process, and the blue, pink, green lines describe recoil energy of nucleus from $3\to2$ process with $\xi=0,1,1.9$, respectively. Same color labeled but instead  shaded bands describe the electron recoil energy from $2\to2$ and $3\to2$ processes. Here we take into account all of the $(n,l)$ xenon electron shells, thus there are 11 discrete points to determine the width of band at each $m_\chi$. The horizontal gray dashed lines represent the detector threshold of current xenon target experiments~\cite{LUX:2016ggv, PandaX-II:2017hlx, XENON:2018voc}. We conclude that the required DM mass in $3\to2$ inelastic scattering is much smaller than that in $2\to2$ elastic process at same recoil energy.

For $3\to 2 $ process, the event rate per unit time per unit energy per unit detector for a target to recoil with energy $E_R$  can be written as
\begin{eqnarray}
{d R_{3\to2}\over dE_R}  =N_T  \int d^3 v_1 d^3 v_2 n_\chi^2 v_1 v_2  { d\sigma^{}_{3\to 2}\over d E_R^{} }, \label{master}
\end{eqnarray}
where $N_T$ is the number of target per unit detector mass, $n_\chi$ is the DM density, $v_1 $ and $v_2 $ are DM velocities in the lab frame.   $d\sigma_{3\to 2}/dE_R$ is the differential cross section. The DM number density can be written as $n_\chi =f(E,v)\rho_\chi/m_\chi$, where $f(E, v)$ is the DM velocity distribution function with $\int d^3 v f(v) =1$ and $\rho_\chi=0.4 {\rm GeV/cm^3}$ being the local DM energy density~\cite{Navarro:1996gj}.  Considering the fact that the recoil energy in Eq.(\ref{recoiler}) as well as the scattering amplitude are almost independent of the DM velocity,  and assuming  both DM particles and target are unpolarized, one can absorb the factor $v_1v_2$ in Eq.(\ref{master}) into cross section and integrate over velocities, resulting  in a characteristic quantity,  $\sigma _{3\to 2} v^2 $.  Typically for the DM-electron scattering, the differential ionization rate is obtained by summing over electron from all possible $(n, l)$ shells,
\begin{align}
\frac{d R_{3\to2}}{d E_R}
=\sum_{n, l}\frac{N_{\rm T} \rho^2_\chi \langle\sigma(\textbf{q})v^2\rangle}{4m^2_\chi E_R} | f^{n,l}_{\rm ion}(k', \textbf{q} )|^2, \label{eq:r3to2}
\end{align}
where we define a reference cross section $\langle\sigma(\textbf{q})v^2\rangle\equiv \textbf{q}
\overline{|\mathcal M(\textbf{q}\,)|^2}/(32 \pi m^2_\chi m_e E'_e)$ and $\textbf{q}, E'_e$ denote the transfer momentum and the total energy of electron in final state. $f^{n, l}_{\rm ion}(k', \textbf{q})$ is the ionization form factor which indicates an electron with initial state $(n,l)$ shell and final state with momentum $k' = \sqrt{2m_e E_R}$.  The ionization form factor is calculated by using the Roothaan- Hartree-Fock radial wave-function for initial electron state and applying plane wave approximation for final state( see Refs.~\cite{Cao:2020bwd,Kopp:2009et} for detail).


\begin{figure}[t]
  \centering
  \includegraphics[width=0.49\textwidth]{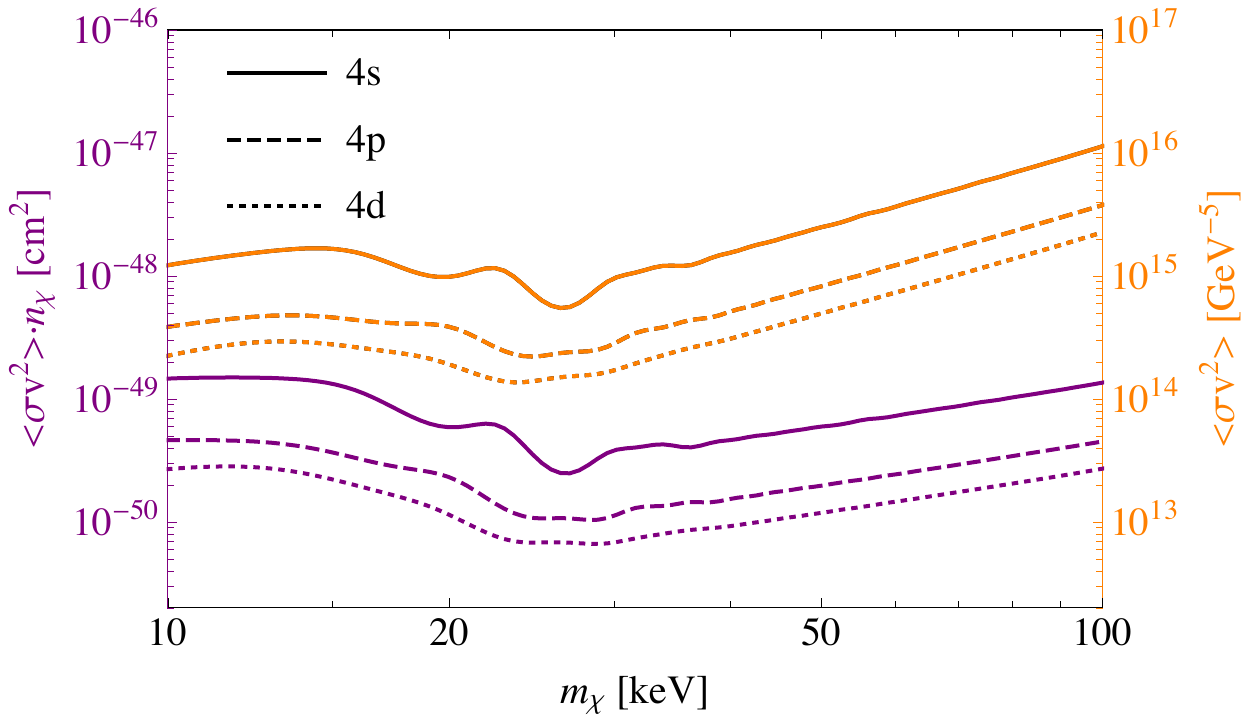}
\caption{Projected limits $\langle \sigma v^2 \rangle$$\cdot$$n_\chi$ (purple with left y-axis) and $\langle \sigma v^2 \rangle$(orange with right y-axis) as a function of $m_\chi$ with $(4s^2, 4p^6, 4d^{10})$ xenon shell for the XENON1T experiment.
}\label{fig:mchi_sigmavv}
\end{figure}

\begin{table}[]
\begin{center}
\begin{tabular}{|c||c|c|c|c|c|c|}
\hline
 (n, l) shell             & $5p^6$ & $5s^2$ & $4d^{10}$ & $4p^6$ & $4s^2$    & $3d^{10}$ \\ \hline
 $|E^{nl}_B|$ [eV]  &12.4      & 25.7      & 75.6          & 163.5    &  213.8     & 710.7        \\ \hline
 \hline
 (n, l) shell             & $3p^6$ & $3s^2$ & $2p^6$     & $2s^2$ & $1s^2$     &               \\ \hline
 $|E^{nl}_B|$ [eV]  & 958.4    & 1093.2  & 4837.7     & 5152.2  & 33317.6   &              \\ \hline
\end{tabular}
\caption{Binding energy for xenon electron shells~\cite{Bunge:1993jsz}.
\vspace{-6mm}}
\label{table:bindenergy}
\end{center}
\end{table}%

In Fig.~\ref{fig:mchi_sigmavv}, we project  sensitivities of XENON1T to the ionization rates given in Eq.~\ref{eq:r3to2}. We calculate the projected constraints on $\langle \sigma v^2 \rangle$$\cdot$$n_\chi~{\rm  (cm^2)}$ and $\langle \sigma v^2 \rangle~({\rm GeV^{-5}})$ for various $m_\chi$ over XENON1T’s full exposure.
As illustrations, we only consider $n=4$ shell electron with $(n, l)=(4s^2, 4p^6, 4d^{10})$, where  the $4d^{10}$ electron is found to be dominate among all shell electron in this mass region. We use the same efficiency as that in the Ref.~\cite{Aprile:2020tmw} when calculating ionization rates. 
We see that the exclusion limit induced by the $3\to 2$ process is significant, which can be further applied to constrain parameter space of specific DM model. Furthermore, fluctuations of curves at low DM regime is induced by the electron binding energy, whose impact turns to be important at low energy transfer.

\section{ The Model}

\begin{figure}[t]
\centering
\subfigure[]{
\includegraphics[width=3.5cm]{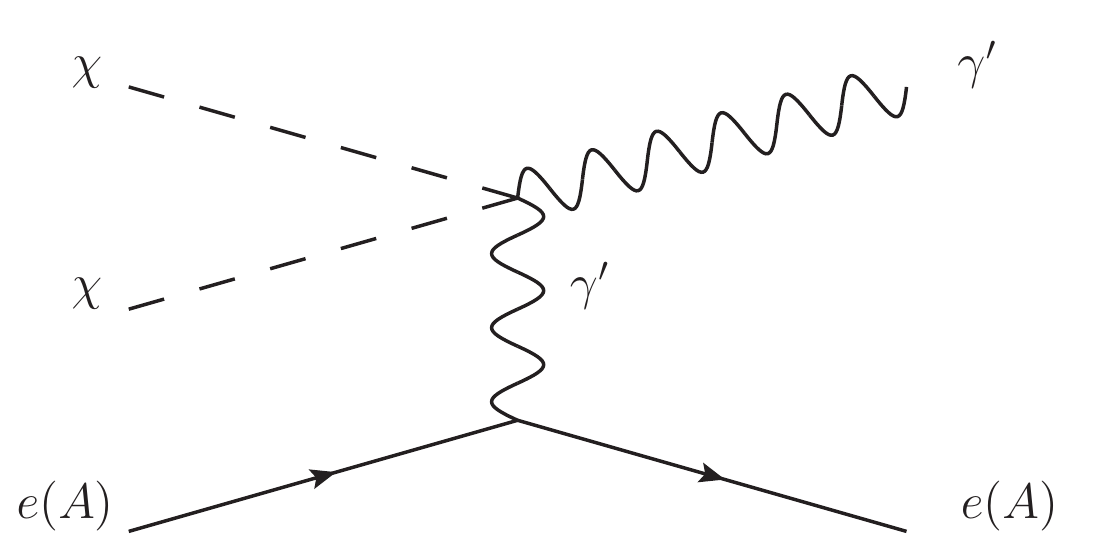}
\label{subfig1:2}
}
\quad
\subfigure[]{
\includegraphics[width=3.5cm]{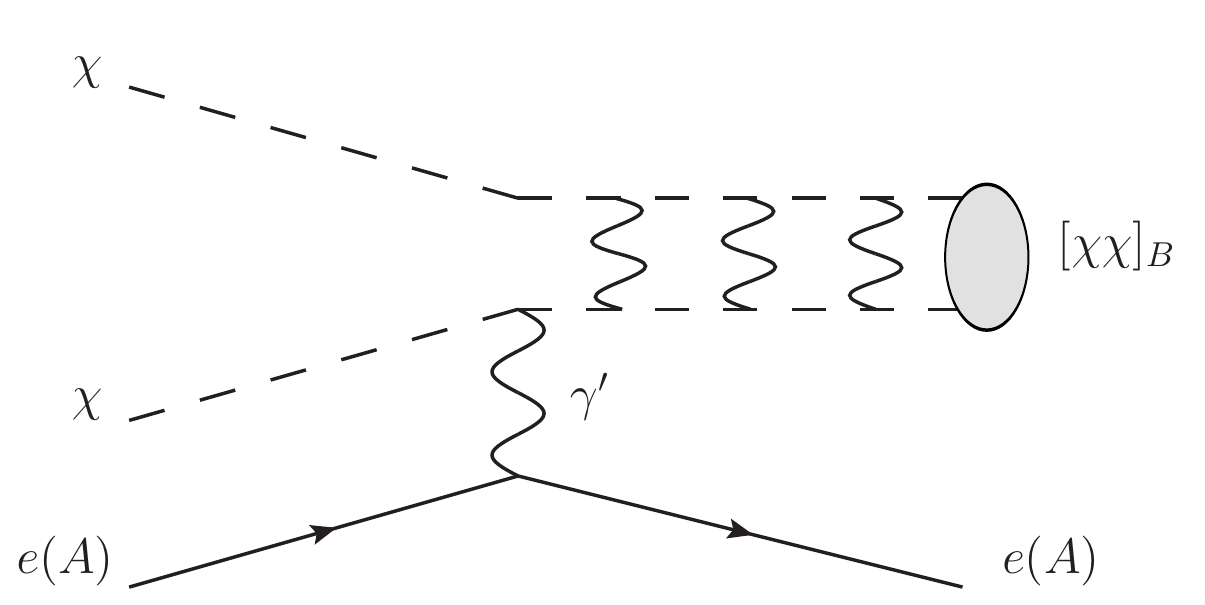}
\label{subfig1:4}
}
\caption{Feynman diagrams of $\chi + \chi + {\rm T} \to \eta + {\rm T}$ scattering process for scalar DM, where $\eta$ indicates one of the final states: dark radiation($\gamma'$) or DM bound state($[\chi\chi]_B$). The left-panel shows real scalar DM with four vertex and the right-panel shows complex scalar DM bound state. }\label{fig:3to2}
\end{figure}

As an illustration, we study the direct detection signal of a complex scalar DM, $\Phi =\frac{1}{\sqrt{2}}(\chi+i\zeta)$, with additional gauge interaction $U(1)_D$, whose gauge field couples to the SM via  the kinetic mixing with photon. Generally there is mass splitting between $\chi$ and $\zeta$ arising from radiation corrections~\cite{Cirelli:2005uq}, leaving the lighter component as the DM candidate.   Here we take $\chi$ as the DM (complex $\Phi$ can be DM candidate for negligible mass splitting). Relevant Lagrangian can be written as
\bea
\mathcal{L} &\supset& (D_\mu \Phi_I)^\dagger (D^\mu \Phi_I) + \frac{m^2_{A'} (A'^\mu)^2}{2} + \epsilon e A'_\mu J^\mu_{\rm em} \label{eq:larsdm} , \,
\eea
where $D_\mu=\partial_\mu -i g_D A_\mu^\prime$ is the covariant  derivative with the new gauge coupling $g_D$, $m_{A^\prime}$ is the mass of new gauge field, $\epsilon$ is the mixing parameter, $J_{\rm em}^{\mu }$ is the electromagnetic current. Due to the mass splitting, the $2\to 2$ scattering process, $\chi +{\rm SM} \to \chi + {\rm SM}$, is usually kinematically forbidden at the tree-level, which is similar to the case of the inert DM~\cite{LopezHonorez:2006gr}, and one-loop corrections to this process are suppressed, leaving the $3\to 2 $ scattering process one possible way out. Two initial DMs scatter on targets via a $3\to2$ process, $\chi + \chi + {\rm T} \to \eta + {\rm T}$, where $\eta$ indicates one of the following final states: $\gamma'$ or $[\chi\chi]_B$, and ${\rm T}$ denotes the target: electron or nuclei.  Relevant Feynman diagrams are shown in Fig.(\ref{fig:3to2}). In the following, we will study the signal of these processes separately.

\textbf{\textit{Scattering into DM bound state via the $3\to 2$ process.}} If the dark photon mass is massless or much lighter than the DM, there is long-range interaction which typically implies the existence of  DM bound states. We study the DM bound states in the non-relativistic regime with non-confining interactions. Ref.~\cite{Petraki:2015hla,Petraki:2016cnz} shows that at lowest order in the coupling and in the non-relativistic regime, the bound state formation cross sections do ${\it not}$ depend on the spin configuration of the DM, therefore we consider  the complex scalar DM case in bound state.
As shown in the right-panel of the Fig.~\ref{fig:3to2}, DMs may form bound state  through the inelastic scattering process. The scattering amplitude can be written as
\begin{align}\label{amp:general}
&i\mathcal M(\chi_1\chi_2 {\rm T}\to [\chi_1\chi_2]_B {\rm T})=\sqrt{Z_\varphi(P_\varphi)}\times
\nonumber\\
&\int\frac{d^4p}{(2\pi)^4}\frac{d^4k}{(2\pi)^4}
\tilde{\Psi}^*_{P_n}(p)\tilde{\Phi}_{K_k}(k)
\mathcal A^{(5)}\tilde{S}_\varphi(P_\varphi) \times i\mathcal M_{\rm SM},
\end{align}
where $Z_\varphi(P_\varphi)$ is the field-strength renormalisation parameter for
the mediator $\varphi$, $\Phi_{K,k}(k)$ and $\Psi_{P,n}(p)$ are the Bethe-Salpeter wavefunctions of DM in initial
scattering states and final bound states respectively. $\mathcal A^{(5)}(P_\varphi,p,k)$ is the 5-point
correlation function and $\tilde{S}_\varphi(P_\varphi)$ is the propagator of $\varphi$.
$i\mathcal M_{\rm SM}$ is the SM part amplitude.
Note that the matrix element squared may depend on the scattering angle $\theta$. Combining the general expression in Eq.(\ref{amp:general}) and the concrete interaction given by Eq.(\ref{eq:larsdm}), the amplitude of $\chi \chi e \to[\chi\chi]_B e$ process at the leading-order is derived in the Supplemental Material.

Assuming the initial two DM has same mass, the energy transferred to the target particle can be expressed as
\begin{align}\label{DeltaE}
\Delta E=2m_{\chi}-\sqrt{\textbf{q}^2+M_{B\{n,l,m\}}^2},
\end{align}
where $M_{B\{n,l,m\}}$ is the mass of final bound state at the $\{n,l,m\}$ level.
For the capture in the final states $\{n,l,m\}$, the binding energy is\cite{Petraki:2015hla}
\begin{align}
\varepsilon_n=-\frac{\mu\alpha^2_D}{2n^2},
\end{align}
where $\alpha_D=\frac{g^2_D}{4\pi}$, $\mu=\frac{1}{2}m_\chi$ is the reduced mass of the dark matter pair. In order to ensure the sufficient efficiency of bound-state formation, we only consider the maximal value for $\varepsilon_n$ with $n=1$. For the ground state $\{100\}$, the total energy of the bound state is $E_{B\{100\}}=\sqrt{\textbf{q}^2+M_{B\{100\}}^2}$ with its mass $M_{B\{100\}}=2m_\chi+\varepsilon_1$~\cite{Petraki:2015hla}. So the transferred energy is roughly $|\varepsilon_1|$, which means that a larger binding energy results in larger recoil energy.

\textbf{\textit{DM - electron scattering in 3$\to$2 process.}}
\label{sec:DM_e}
In this section, we consider the real scalar DM-electron  inelastic scattering via the $3\to2$ process. We assume that the mass of dark photon is ultralight($m_{A'} \ll m_\chi$), then the recoil energy for the target electron is,
\be
E_R \simeq 2m_\chi-q-|E^e_B|,
\label{eq:q_e}
\ee
where $E^e_B$ is the binding energy of electron in atom and $q$ is the energy of dark photon. Substituting Eq.~\ref{eq:q_e} and the squared matrix element given in supplemental material into Eq.~\ref{eq:r3to2}, one obtains the whole expression for the rate of DM scattering off the electron. 

\begin{figure}[t]
  \centering
  \includegraphics[width=0.45\textwidth]{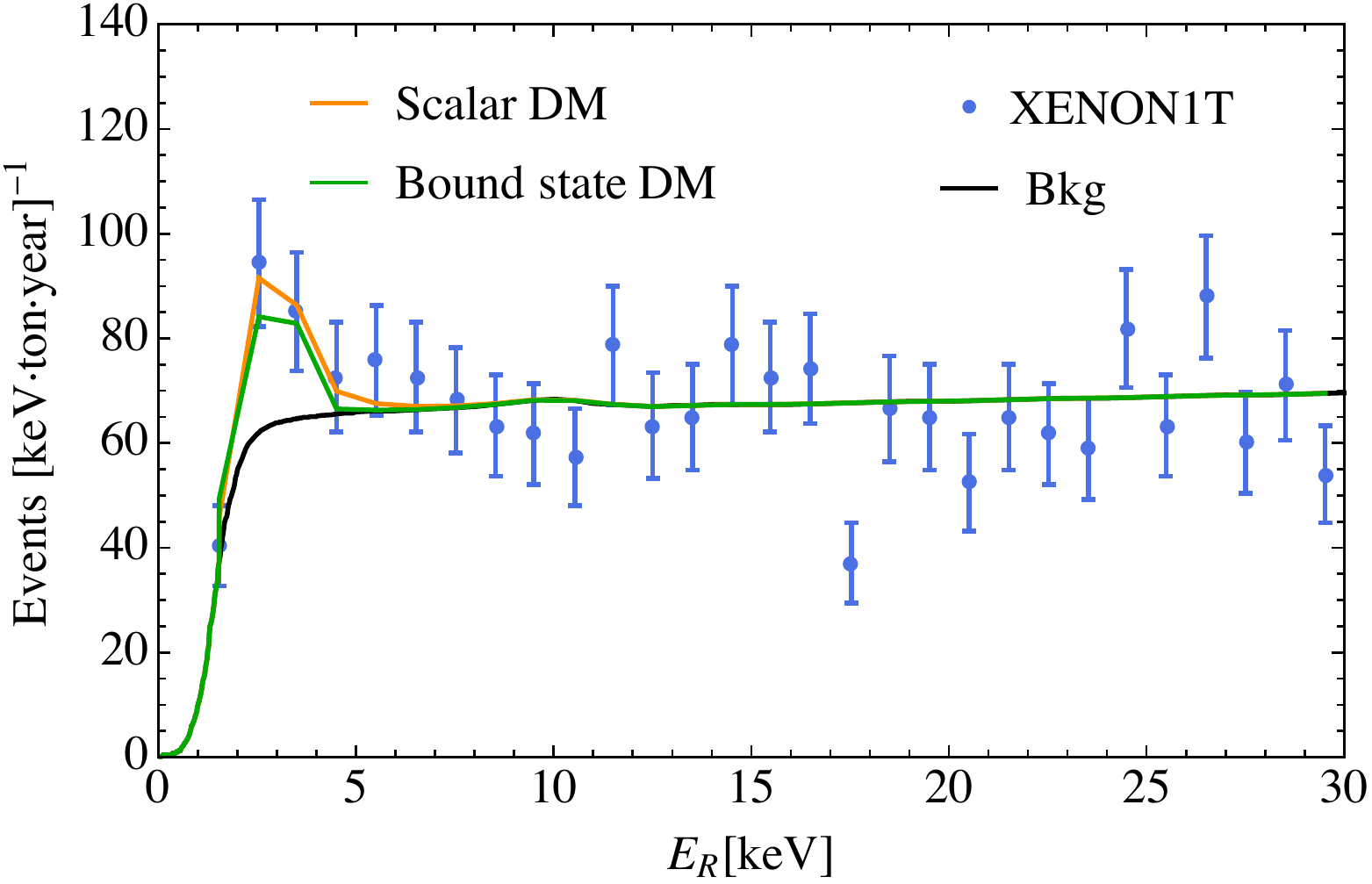}
\caption{The $3\to2$ event rates in XENON1T excess for the scalar and bound state DM. The colored solid lines denote the signal plus background and black line is the background only event rates. The blue points is the XENON1t measured data.
 The best fit parameters of scalar(bound state) DM: $m_\chi$ = 29.7(96.6) keV, $\epsilon=2\times 10^{-6}(2.2\times 10^{-6})$. The ratio between the DM and dark photon is $\xi=10^{-3}$ and the gauge coupling is $|g_D|=1(3)$ for scalar(bound state) DM.
}\label{fig:rate3to2}
\end{figure}

In Fig.~\ref{fig:rate3to2}, we show the $3\to2$ event rates  for the scalar DM and bound state DM together with the best fit points for the XENON1T excess~\cite{Aprile:2020tmw}. To perform the sensitivity analysis, we consider a simple chi-squared $\chi^2$ test. The statistical significance can be identified by $\delta \chi^2= \chi^2_{\rm min}-\chi^2_{\rm bkg}$, where $\chi^2_{\rm bkg}$=45.5 for the background obtained by the XENON Collaboration~\cite{Aprile:2020tmw}. The best fit parameters are $m_\chi$ = 29.7 keV, $\epsilon=2\times 10^{-6}$ for scalar DM, $m_\chi$=96.6 keV, $\epsilon=2.2\times 10^{-6}$ for bound state DM, which corresponds to $\chi^2$= 34.8 and 36.9, respectively. The mass ratio between dark photon and  DM is setting as $\xi=10^{-3}$ and the gauge coupling is $|g_D|=1(3)$ for scalar (bound state) DM.
The signal from $3\to2$ contribution is denoted by orange and green solid lines, while the background prediction from XENON1T is shown with black line and the blue points is the experimental data of XENON1T.


\begin{figure}[t]
  \centering
  \includegraphics[width=0.45\textwidth]{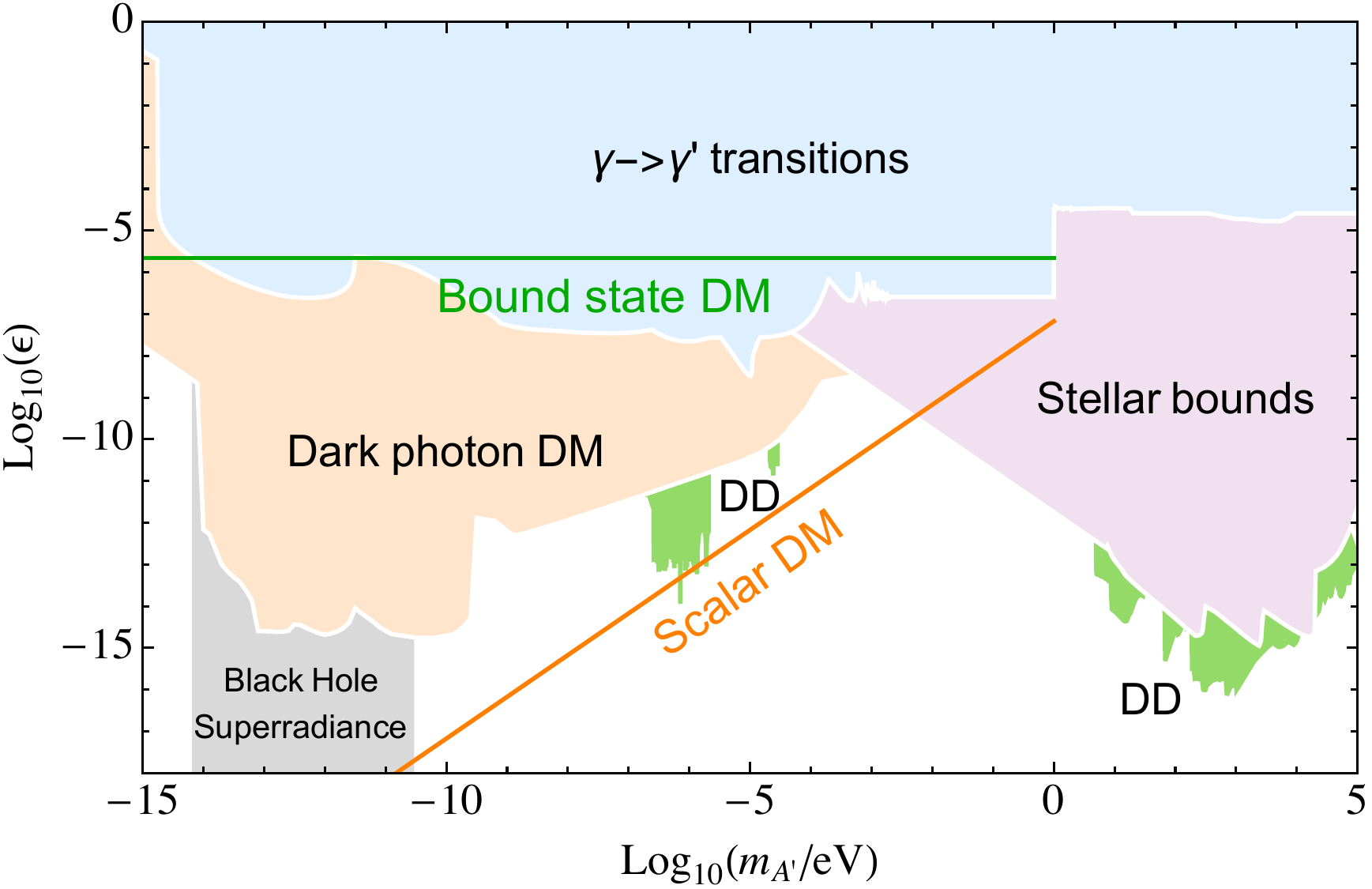}
\caption{The current constraint on dark photon mass($m_{A'}$) and mixing parameter($\epsilon$). The shaded regions are divide into three categories based on cosmological(orange), experimental(blue, green), and astrophysical(purple) bounds. The gray band indicates the mass window from black holes superradiance. DD denotes constraint from the direct detection experiment. Same labels as the Fig.~\ref{fig:rate3to2}, the solid lines denote the value of $\epsilon$ required to explain the XENON1T excess in term of the $m_{A'}$ in $3\to2$ inelastic scattering, where the dark photon mass $m_{A'}\le $1 eV.
}\label{fig:darkphotonlimits}
\end{figure}

We further present the current constraint on dark photon mass($m_{A'}$) and  the mixing parameter($\epsilon$) in Fig.~\ref{fig:darkphotonlimits}. The shaded region including various limits can be divide into three categories based on cosmological(orange), experimental(blue, green), and astrophysical(purple) bounds. The gray band indicates the mass window from black holes superradiance. A detailed description of each bound can be found in a recent review~\cite{Caputo:2021eaa} and references therein. We show the value of $\epsilon$ required to explain the XENON1T excess in term of the $m_{A'}$ with the best fit parameters in $3\to2$ inelastic scattering. As shown in Fig.~\ref{fig:darkphotonlimits}, the mixing parameter $\epsilon$ for bound state DM is independent of $m_{A'}$ when the dark photon is ultralight($m_{A'}\le $1 eV), which is already excluded by the current bounds. However, the scalar DM indicates that the $\epsilon$ is proportional to $m_{A'}$, which leaves a substantial of unconstrained region.

\textbf{\textit{DM - nucleus scattering via 3$\to$2 process.}}
\label{sec:DM_n}
DM-nuclei scattering has been a promising channel of DM direct detections and it already put strong constraint on WIMP-nuclei interactions. However this constraint dramatically weakens for DM mass smaller than about 1 GeV. This is due to the rapidly decreasing sensitivity  at low recoil energies.  The traditional DM-nuclei scattering leaves light DM candidates poorly explored by direct searches. Several approaches have been proposed to directly detect DM at this mass range, such as inelastic DM-nuclei scattering~\cite{Giudice:2017zke} and boosted DM, where a fraction of DM gets a high velocity due to a number of different mechanisms~\cite{Agashe:2014yua}.  The model independent constraint on the DM-nucleon cross section is about $10^{-31}~{\rm cm^2}$ for cosmic ray boosted DM.  In this section we consider the possibility of  detecting Sub-GeV DM via the $3\to 2$ process.
Taking scalar DM as an example, the Feynman diagram is given in the left-panel of the Fig.~(\ref{fig:3to2}) and the $\sigma^{\chi N}_{3\to 2} v^2$ can be written as
\begin{eqnarray}\label{eq:sigmaN}
\sigma^{\chi N}_{3\to 2} v^2  \approx {	q_0\over 32 \pi m_\chi^2 m_N^2} \left|{\cal M}\right|^2 *C_N^2,
\end{eqnarray}
where $|{\cal M}|^2$ is the squared matrix element whose expression is exactly given in supplement material up to replacement $m_e \leftrightarrow m_N$,   $C_N^{} =2 c_u^p +c_d^p$ being the matching factor from the quark level to nucleon level, $q_0\simeq\sqrt{4-\xi^2}m_\chi$ being the momentum transfer.

\begin{figure}[t]
  \centering
  \includegraphics[width=0.45\textwidth]{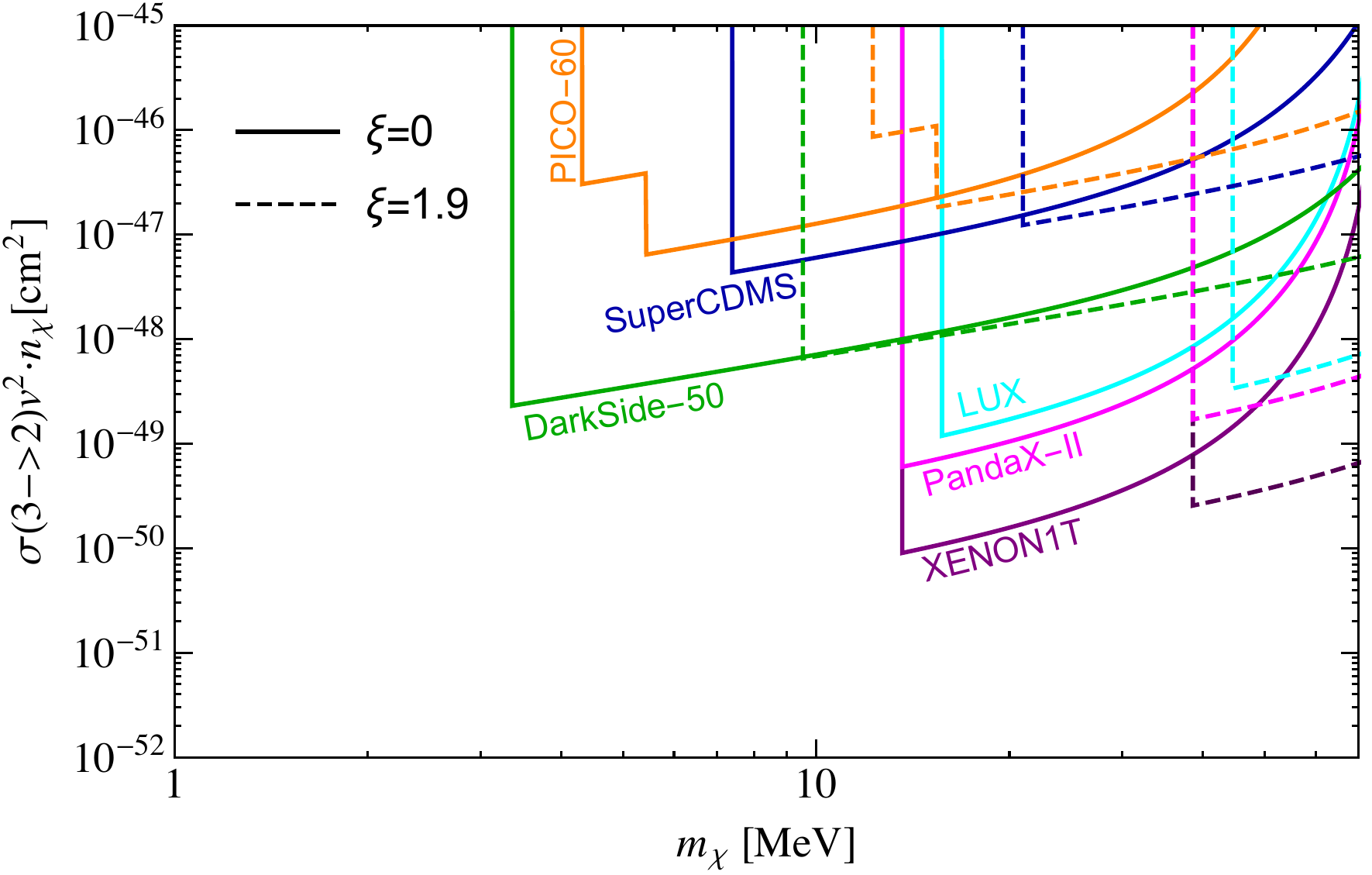}
\caption{Projected limits of several current experiments to $\sigma_{3\to2}v^2$$\cdot$$n_\chi $ vs. $m_\chi$. The solid and dashed lines denote $\xi=0, 1.9$, respectively, corresponding to dark photon and bound state final state.}
\label{fig:mchisigma3to2}
\end{figure}

Refs.~\cite{Dror:2019dib,Dror:2019onn} have studied the absorption of a Fermionic DM in direct detection, where the energy transfer equals to the dark matter mass. The total rate for $3\to2$ inelastic scattering process is similar to that in the Refs.~\cite{Dror:2019dib,Dror:2019onn} if the mass of dark photon in the final state  is massless or ultralight. The total rate  with multiple nuclei can be written as,
\be
R=\left(\frac{\rho_\chi}{m_\chi}\right)^2  \sigma^{\chi N}_{\rm 3\to2} v^2 \sum_i N_{T,i} Z^2_i F^2_i \Theta(E^0_{R,i}-E_{\rm th}),
\ee
where $\sigma^{\chi N}_{3\to2}$ is the $3\to2$ inelastic cross section per nucleon, $Z_i$ is the atomic number for $i$th target nuclei, $F$ is the Helm form factor~\cite{Lewin:1995rx}. The recoil energy of nuclei is monoenergetic and the signature is peaked at $E_R=E^0_R \simeq (4-\xi^2)m^2_\chi/(2m_A)$. $E_{\rm th}$ is the threshold of direct detection. Fig.~\ref{fig:mchisigma3to2} shows the projected limits of current experiments to $\sigma^{\chi N}_{3\to2} \cdot n_\chi $ as a function of $m_\chi$, including LUX~\cite{LUX:2016ggv}, PandaX-II~\cite{PandaX-II:2017hlx}, XENON1T~\cite{XENON:2018voc}, PICO-60(${\rm C_3F_8}$)~\cite{PICO:2017tgi}, SuperCDMS~\cite{SuperCDMS:2014cds}, DarkSide-50~\cite{DarkSide:2018bpj}. The solid and dashed lines denote $\xi=0, 1.9$, which correspond to  scattering into dark photon and bound state final state, respectively.


\section{conclusion}

Direct detections of DM in underground laboratories have been a promising way of exploring the particle nature of DM. Given the situation that all searches of WIMPs have turned up null even for exponentially  increased exposure, direct detection of sub-GeV DM becomes more and more important, whose exclusion limit is still very high due to the limitation of the detector threshold. In this Letter, we have proposed a new direct detection strategy  via  $3-$body inelastic scattering process.  The energy transfer to detector target via this process is enhanced  compared with the $2\to 2$ scattering process. So it can be applied to the direct detection of lighter DM.  We have extracted the generic  physical observable for this process and presented its effects in the direct detection of  complex scalar DM. It should be mentioned that this method  is also applicable to the direct detection of fermion DM.

\begin{acknowledgments}
This work was supported by the National Natural Science Foundation of China under grant No. 11775025 and No. 12175027.
\end{acknowledgments}

\bibliography{refs}

\begin{thebibliography}{64}%
\makeatletter
\providecommand \@ifxundefined [1]{%
 \@ifx{#1\undefined}
}%
\providecommand \@ifnum [1]{%
 \ifnum #1\expandafter \@firstoftwo
 \else \expandafter \@secondoftwo
 \fi
}%
\providecommand \@ifx [1]{%
 \ifx #1\expandafter \@firstoftwo
 \else \expandafter \@secondoftwo
 \fi
}%
\providecommand \natexlab [1]{#1}%
\providecommand \enquote  [1]{``#1''}%
\providecommand \bibnamefont  [1]{#1}%
\providecommand \bibfnamefont [1]{#1}%
\providecommand \citenamefont [1]{#1}%
\providecommand \href@noop [0]{\@secondoftwo}%
\providecommand \href [0]{\begingroup \@sanitize@url \@href}%
\providecommand \@href[1]{\@@startlink{#1}\@@href}%
\providecommand \@@href[1]{\endgroup#1\@@endlink}%
\providecommand \@sanitize@url [0]{\catcode `\\12\catcode `\$12\catcode
  `\&12\catcode `\#12\catcode `\^12\catcode `\_12\catcode `\%12\relax}%
\providecommand \@@startlink[1]{}%
\providecommand \@@endlink[0]{}%
\providecommand \url  [0]{\begingroup\@sanitize@url \@url }%
\providecommand \@url [1]{\endgroup\@href {#1}{\urlprefix }}%
\providecommand \urlprefix  [0]{URL }%
\providecommand \Eprint [0]{\href }%
\providecommand \doibase [0]{http://dx.doi.org/}%
\providecommand \selectlanguage [0]{\@gobble}%
\providecommand \bibinfo  [0]{\@secondoftwo}%
\providecommand \bibfield  [0]{\@secondoftwo}%
\providecommand \translation [1]{[#1]}%
\providecommand \BibitemOpen [0]{}%
\providecommand \bibitemStop [0]{}%
\providecommand \bibitemNoStop [0]{.\EOS\space}%
\providecommand \EOS [0]{\spacefactor3000\relax}%
\providecommand \BibitemShut  [1]{\csname bibitem#1\endcsname}%
\let\auto@bib@innerbib\@empty
\bibitem [{\citenamefont {Liu}\ \emph {et~al.}(2017)\citenamefont {Liu},
  \citenamefont {Chen},\ and\ \citenamefont {Ji}}]{Liu:2017drf}%
  \BibitemOpen
  \bibfield  {author} {\bibinfo {author} {\bibfnamefont {J.}~\bibnamefont
  {Liu}}, \bibinfo {author} {\bibfnamefont {X.}~\bibnamefont {Chen}}, \ and\
  \bibinfo {author} {\bibfnamefont {X.}~\bibnamefont {Ji}},\ }\href {\doibase
  10.1038/nphys4039} {\bibfield  {journal} {\bibinfo  {journal} {Nature Phys.}\
  }\textbf {\bibinfo {volume} {13}},\ \bibinfo {pages} {212} (\bibinfo {year}
  {2017})},\ \Eprint {http://arxiv.org/abs/1709.00688} {arXiv:1709.00688
  [astro-ph.CO]} \BibitemShut {NoStop}%
\bibitem [{\citenamefont {Lin}(2019)}]{Lin:2019uvt}%
  \BibitemOpen
  \bibfield  {author} {\bibinfo {author} {\bibfnamefont {T.}~\bibnamefont
  {Lin}},\ }\href {\doibase 10.22323/1.333.0009} {\bibfield  {journal}
  {\bibinfo  {journal} {PoS}\ }\textbf {\bibinfo {volume} {333}},\ \bibinfo
  {pages} {009} (\bibinfo {year} {2019})},\ \Eprint
  {http://arxiv.org/abs/1904.07915} {arXiv:1904.07915 [hep-ph]} \BibitemShut
  {NoStop}%
\bibitem [{\citenamefont {Griest}\ and\ \citenamefont
  {Kamionkowski}(1990)}]{Griest:1989wd}%
  \BibitemOpen
  \bibfield  {author} {\bibinfo {author} {\bibfnamefont {K.}~\bibnamefont
  {Griest}}\ and\ \bibinfo {author} {\bibfnamefont {M.}~\bibnamefont
  {Kamionkowski}},\ }\href {\doibase 10.1103/PhysRevLett.64.615} {\bibfield
  {journal} {\bibinfo  {journal} {Phys. Rev. Lett.}\ }\textbf {\bibinfo
  {volume} {64}},\ \bibinfo {pages} {615} (\bibinfo {year} {1990})}\BibitemShut
  {NoStop}%
\bibitem [{\citenamefont {Steigman}\ and\ \citenamefont
  {Turner}(1985)}]{Steigman:1984ac}%
  \BibitemOpen
  \bibfield  {author} {\bibinfo {author} {\bibfnamefont {G.}~\bibnamefont
  {Steigman}}\ and\ \bibinfo {author} {\bibfnamefont {M.~S.}\ \bibnamefont
  {Turner}},\ }\href {\doibase 10.1016/0550-3213(85)90537-1} {\bibfield
  {journal} {\bibinfo  {journal} {Nucl. Phys. B}\ }\textbf {\bibinfo {volume}
  {253}},\ \bibinfo {pages} {375} (\bibinfo {year} {1985})}\BibitemShut
  {NoStop}%
\bibitem [{\citenamefont {Aprile}\ \emph
  {et~al.}(2018{\natexlab{a}})\citenamefont {Aprile} \emph
  {et~al.}}]{XENON:2018bec}%
  \BibitemOpen
  \bibfield  {author} {\bibinfo {author} {\bibfnamefont {E.}~\bibnamefont
  {Aprile}} \emph {et~al.} (\bibinfo {collaboration} {XENON}),\ }\href
  {\doibase 10.1103/PhysRevLett.121.111302} {\bibfield  {journal} {\bibinfo
  {journal} {Phys. Rev. Lett.}\ }\textbf {\bibinfo {volume} {121}},\ \bibinfo
  {pages} {111302} (\bibinfo {year} {2018}{\natexlab{a}})},\ \Eprint
  {http://arxiv.org/abs/1805.12562} {arXiv:1805.12562 [astro-ph.CO]}
  \BibitemShut {NoStop}%
\bibitem [{\citenamefont {Cui}\ \emph {et~al.}(2017{\natexlab{a}})\citenamefont
  {Cui} \emph {et~al.}}]{PandaX-II:2017jmq}%
  \BibitemOpen
  \bibfield  {author} {\bibinfo {author} {\bibfnamefont {X.}~\bibnamefont
  {Cui}} \emph {et~al.} (\bibinfo {collaboration} {PandaX-II}),\ }\href
  {\doibase 10.1103/PhysRevLett.119.181302} {\bibfield  {journal} {\bibinfo
  {journal} {Phys. Rev. Lett.}\ }\textbf {\bibinfo {volume} {119}},\ \bibinfo
  {pages} {181302} (\bibinfo {year} {2017}{\natexlab{a}})},\ \Eprint
  {http://arxiv.org/abs/1708.06917} {arXiv:1708.06917 [astro-ph.CO]}
  \BibitemShut {NoStop}%
\bibitem [{\citenamefont {Chao}\ \emph {et~al.}(2021)\citenamefont {Chao},
  \citenamefont {Li},\ and\ \citenamefont {Liao}}]{Chao:2021vja}%
  \BibitemOpen
  \bibfield  {author} {\bibinfo {author} {\bibfnamefont {W.}~\bibnamefont
  {Chao}}, \bibinfo {author} {\bibfnamefont {T.}~\bibnamefont {Li}}, \ and\
  \bibinfo {author} {\bibfnamefont {J.}~\bibnamefont {Liao}},\ }\href@noop {}
  {\  (\bibinfo {year} {2021})},\ \Eprint {http://arxiv.org/abs/2108.05608}
  {arXiv:2108.05608 [hep-ph]} \BibitemShut {NoStop}%
\bibitem [{\citenamefont {Das}\ and\ \citenamefont {Sen}(2021)}]{Das:2021lcr}%
  \BibitemOpen
  \bibfield  {author} {\bibinfo {author} {\bibfnamefont {A.}~\bibnamefont
  {Das}}\ and\ \bibinfo {author} {\bibfnamefont {M.}~\bibnamefont {Sen}},\
  }\href@noop {} {\  (\bibinfo {year} {2021})},\ \Eprint
  {http://arxiv.org/abs/2104.00027} {arXiv:2104.00027 [hep-ph]} \BibitemShut
  {NoStop}%
\bibitem [{\citenamefont {Jho}\ \emph {et~al.}(2021)\citenamefont {Jho},
  \citenamefont {Park}, \citenamefont {Park},\ and\ \citenamefont
  {Tseng}}]{Jho:2021rmn}%
  \BibitemOpen
  \bibfield  {author} {\bibinfo {author} {\bibfnamefont {Y.}~\bibnamefont
  {Jho}}, \bibinfo {author} {\bibfnamefont {J.-C.}\ \bibnamefont {Park}},
  \bibinfo {author} {\bibfnamefont {S.~C.}\ \bibnamefont {Park}}, \ and\
  \bibinfo {author} {\bibfnamefont {P.-Y.}\ \bibnamefont {Tseng}},\ }\href@noop
  {} {\  (\bibinfo {year} {2021})},\ \Eprint {http://arxiv.org/abs/2101.11262}
  {arXiv:2101.11262 [hep-ph]} \BibitemShut {NoStop}%
\bibitem [{\citenamefont {Su}\ \emph {et~al.}(2020)\citenamefont {Su},
  \citenamefont {Wang}, \citenamefont {Wu}, \citenamefont {Yang},\ and\
  \citenamefont {Zhu}}]{Su:2020zny}%
  \BibitemOpen
  \bibfield  {author} {\bibinfo {author} {\bibfnamefont {L.}~\bibnamefont
  {Su}}, \bibinfo {author} {\bibfnamefont {W.}~\bibnamefont {Wang}}, \bibinfo
  {author} {\bibfnamefont {L.}~\bibnamefont {Wu}}, \bibinfo {author}
  {\bibfnamefont {J.~M.}\ \bibnamefont {Yang}}, \ and\ \bibinfo {author}
  {\bibfnamefont {B.}~\bibnamefont {Zhu}},\ }\href {\doibase
  10.1103/PhysRevD.102.115028} {\bibfield  {journal} {\bibinfo  {journal}
  {Phys. Rev. D}\ }\textbf {\bibinfo {volume} {102}},\ \bibinfo {pages}
  {115028} (\bibinfo {year} {2020})},\ \Eprint
  {http://arxiv.org/abs/2006.11837} {arXiv:2006.11837 [hep-ph]} \BibitemShut
  {NoStop}%
\bibitem [{\citenamefont {Fornal}\ \emph {et~al.}(2020)\citenamefont {Fornal},
  \citenamefont {Sandick}, \citenamefont {Shu}, \citenamefont {Su},\ and\
  \citenamefont {Zhao}}]{Fornal:2020npv}%
  \BibitemOpen
  \bibfield  {author} {\bibinfo {author} {\bibfnamefont {B.}~\bibnamefont
  {Fornal}}, \bibinfo {author} {\bibfnamefont {P.}~\bibnamefont {Sandick}},
  \bibinfo {author} {\bibfnamefont {J.}~\bibnamefont {Shu}}, \bibinfo {author}
  {\bibfnamefont {M.}~\bibnamefont {Su}}, \ and\ \bibinfo {author}
  {\bibfnamefont {Y.}~\bibnamefont {Zhao}},\ }\href {\doibase
  10.1103/PhysRevLett.125.161804} {\bibfield  {journal} {\bibinfo  {journal}
  {Phys. Rev. Lett.}\ }\textbf {\bibinfo {volume} {125}},\ \bibinfo {pages}
  {161804} (\bibinfo {year} {2020})},\ \Eprint
  {http://arxiv.org/abs/2006.11264} {arXiv:2006.11264 [hep-ph]} \BibitemShut
  {NoStop}%
\bibitem [{\citenamefont {An}\ and\ \citenamefont {Yang}(2021)}]{An:2020tcg}%
  \BibitemOpen
  \bibfield  {author} {\bibinfo {author} {\bibfnamefont {H.}~\bibnamefont
  {An}}\ and\ \bibinfo {author} {\bibfnamefont {D.}~\bibnamefont {Yang}},\
  }\href {\doibase 10.1016/j.physletb.2021.136408} {\bibfield  {journal}
  {\bibinfo  {journal} {Phys. Lett. B}\ }\textbf {\bibinfo {volume} {818}},\
  \bibinfo {pages} {136408} (\bibinfo {year} {2021})},\ \Eprint
  {http://arxiv.org/abs/2006.15672} {arXiv:2006.15672 [hep-ph]} \BibitemShut
  {NoStop}%
\bibitem [{\citenamefont {He}\ \emph {et~al.}(2021)\citenamefont {He},
  \citenamefont {Wang},\ and\ \citenamefont {Zheng}}]{He:2020wjs}%
  \BibitemOpen
  \bibfield  {author} {\bibinfo {author} {\bibfnamefont {H.-J.}\ \bibnamefont
  {He}}, \bibinfo {author} {\bibfnamefont {Y.-C.}\ \bibnamefont {Wang}}, \ and\
  \bibinfo {author} {\bibfnamefont {J.}~\bibnamefont {Zheng}},\ }\href
  {\doibase 10.1088/1475-7516/2021/01/042} {\bibfield  {journal} {\bibinfo
  {journal} {JCAP}\ }\textbf {\bibinfo {volume} {01}},\ \bibinfo {pages} {042}
  (\bibinfo {year} {2021})},\ \Eprint {http://arxiv.org/abs/2007.04963}
  {arXiv:2007.04963 [hep-ph]} \BibitemShut {NoStop}%
\bibitem [{\citenamefont {Baryakhtar}\ \emph {et~al.}(2020)\citenamefont
  {Baryakhtar}, \citenamefont {Berlin}, \citenamefont {Liu},\ and\
  \citenamefont {Weiner}}]{Baryakhtar:2020rwy}%
  \BibitemOpen
  \bibfield  {author} {\bibinfo {author} {\bibfnamefont {M.}~\bibnamefont
  {Baryakhtar}}, \bibinfo {author} {\bibfnamefont {A.}~\bibnamefont {Berlin}},
  \bibinfo {author} {\bibfnamefont {H.}~\bibnamefont {Liu}}, \ and\ \bibinfo
  {author} {\bibfnamefont {N.}~\bibnamefont {Weiner}},\ }\href@noop {} {\
  (\bibinfo {year} {2020})},\ \Eprint {http://arxiv.org/abs/2006.13918}
  {arXiv:2006.13918 [hep-ph]} \BibitemShut {NoStop}%
\bibitem [{\citenamefont {Song}\ \emph {et~al.}(2021)\citenamefont {Song},
  \citenamefont {Nagorny},\ and\ \citenamefont {Vincent}}]{Song:2021yar}%
  \BibitemOpen
  \bibfield  {author} {\bibinfo {author} {\bibfnamefont {N.}~\bibnamefont
  {Song}}, \bibinfo {author} {\bibfnamefont {S.}~\bibnamefont {Nagorny}}, \
  and\ \bibinfo {author} {\bibfnamefont {A.~C.}\ \bibnamefont {Vincent}},\
  }\href@noop {} {\  (\bibinfo {year} {2021})},\ \Eprint
  {http://arxiv.org/abs/2104.09517} {arXiv:2104.09517 [hep-ph]} \BibitemShut
  {NoStop}%
\bibitem [{\citenamefont {Aad}\ \emph {et~al.}(2021)\citenamefont {Aad} \emph
  {et~al.}}]{ATLAS:2021kog}%
  \BibitemOpen
  \bibfield  {author} {\bibinfo {author} {\bibfnamefont {G.}~\bibnamefont
  {Aad}} \emph {et~al.} (\bibinfo {collaboration} {ATLAS}),\ }\href {\doibase
  10.1007/JHEP07(2021)005} {\bibfield  {journal} {\bibinfo  {journal} {JHEP}\
  }\textbf {\bibinfo {volume} {07}},\ \bibinfo {pages} {005} (\bibinfo {year}
  {2021})},\ \Eprint {http://arxiv.org/abs/2103.01918} {arXiv:2103.01918
  [hep-ex]} \BibitemShut {NoStop}%
\bibitem [{\citenamefont {He}\ \emph {et~al.}(2020)\citenamefont {He},
  \citenamefont {Wang},\ and\ \citenamefont {Zheng}}]{He:2020sat}%
  \BibitemOpen
  \bibfield  {author} {\bibinfo {author} {\bibfnamefont {H.-J.}\ \bibnamefont
  {He}}, \bibinfo {author} {\bibfnamefont {Y.-C.}\ \bibnamefont {Wang}}, \ and\
  \bibinfo {author} {\bibfnamefont {J.}~\bibnamefont {Zheng}},\ }\href@noop {}
  {\  (\bibinfo {year} {2020})},\ \Eprint {http://arxiv.org/abs/2012.05891}
  {arXiv:2012.05891 [hep-ph]} \BibitemShut {NoStop}%
\bibitem [{\citenamefont {Harigaya}\ \emph {et~al.}(2020)\citenamefont
  {Harigaya}, \citenamefont {Nakai},\ and\ \citenamefont
  {Suzuki}}]{Harigaya:2020ckz}%
  \BibitemOpen
  \bibfield  {author} {\bibinfo {author} {\bibfnamefont {K.}~\bibnamefont
  {Harigaya}}, \bibinfo {author} {\bibfnamefont {Y.}~\bibnamefont {Nakai}}, \
  and\ \bibinfo {author} {\bibfnamefont {M.}~\bibnamefont {Suzuki}},\ }\href
  {\doibase 10.1016/j.physletb.2020.135729} {\bibfield  {journal} {\bibinfo
  {journal} {Phys. Lett. B}\ }\textbf {\bibinfo {volume} {809}},\ \bibinfo
  {pages} {135729} (\bibinfo {year} {2020})},\ \Eprint
  {http://arxiv.org/abs/2006.11938} {arXiv:2006.11938 [hep-ph]} \BibitemShut
  {NoStop}%
\bibitem [{\citenamefont {Jacobsen}\ \emph {et~al.}(2021)\citenamefont
  {Jacobsen}, \citenamefont {Freese}, \citenamefont {Kelso}, \citenamefont
  {Sandick},\ and\ \citenamefont {Stengel}}]{Jacobsen:2021vbr}%
  \BibitemOpen
  \bibfield  {author} {\bibinfo {author} {\bibfnamefont {S.}~\bibnamefont
  {Jacobsen}}, \bibinfo {author} {\bibfnamefont {K.}~\bibnamefont {Freese}},
  \bibinfo {author} {\bibfnamefont {C.}~\bibnamefont {Kelso}}, \bibinfo
  {author} {\bibfnamefont {P.}~\bibnamefont {Sandick}}, \ and\ \bibinfo
  {author} {\bibfnamefont {P.}~\bibnamefont {Stengel}},\ }\href@noop {} {\
  (\bibinfo {year} {2021})},\ \Eprint {http://arxiv.org/abs/2102.08367}
  {arXiv:2102.08367 [hep-ph]} \BibitemShut {NoStop}%
\bibitem [{\citenamefont {Borah}\ \emph {et~al.}(2020)\citenamefont {Borah},
  \citenamefont {Mahapatra}, \citenamefont {Nanda},\ and\ \citenamefont
  {Sahu}}]{Borah:2020jzi}%
  \BibitemOpen
  \bibfield  {author} {\bibinfo {author} {\bibfnamefont {D.}~\bibnamefont
  {Borah}}, \bibinfo {author} {\bibfnamefont {S.}~\bibnamefont {Mahapatra}},
  \bibinfo {author} {\bibfnamefont {D.}~\bibnamefont {Nanda}}, \ and\ \bibinfo
  {author} {\bibfnamefont {N.}~\bibnamefont {Sahu}},\ }\href {\doibase
  10.1016/j.physletb.2020.135933} {\bibfield  {journal} {\bibinfo  {journal}
  {Phys. Lett. B}\ }\textbf {\bibinfo {volume} {811}},\ \bibinfo {pages}
  {135933} (\bibinfo {year} {2020})},\ \Eprint
  {http://arxiv.org/abs/2007.10754} {arXiv:2007.10754 [hep-ph]} \BibitemShut
  {NoStop}%
\bibitem [{\citenamefont {Chao}\ \emph {et~al.}(2020)\citenamefont {Chao},
  \citenamefont {Gao},\ and\ \citenamefont {Jin}}]{Chao:2020yro}%
  \BibitemOpen
  \bibfield  {author} {\bibinfo {author} {\bibfnamefont {W.}~\bibnamefont
  {Chao}}, \bibinfo {author} {\bibfnamefont {Y.}~\bibnamefont {Gao}}, \ and\
  \bibinfo {author} {\bibfnamefont {M.~j.}\ \bibnamefont {Jin}},\ }\href@noop
  {} {\  (\bibinfo {year} {2020})},\ \Eprint {http://arxiv.org/abs/2006.16145}
  {arXiv:2006.16145 [hep-ph]} \BibitemShut {NoStop}%
\bibitem [{\citenamefont {Dutta}\ \emph {et~al.}(2021)\citenamefont {Dutta},
  \citenamefont {Mahapatra}, \citenamefont {Borah},\ and\ \citenamefont
  {Sahu}}]{Dutta:2021wbn}%
  \BibitemOpen
  \bibfield  {author} {\bibinfo {author} {\bibfnamefont {M.}~\bibnamefont
  {Dutta}}, \bibinfo {author} {\bibfnamefont {S.}~\bibnamefont {Mahapatra}},
  \bibinfo {author} {\bibfnamefont {D.}~\bibnamefont {Borah}}, \ and\ \bibinfo
  {author} {\bibfnamefont {N.}~\bibnamefont {Sahu}},\ }\href {\doibase
  10.1103/PhysRevD.103.095018} {\bibfield  {journal} {\bibinfo  {journal}
  {Phys. Rev. D}\ }\textbf {\bibinfo {volume} {103}},\ \bibinfo {pages}
  {095018} (\bibinfo {year} {2021})},\ \Eprint
  {http://arxiv.org/abs/2101.06472} {arXiv:2101.06472 [hep-ph]} \BibitemShut
  {NoStop}%
\bibitem [{\citenamefont {Keung}\ \emph {et~al.}(2021)\citenamefont {Keung},
  \citenamefont {Marfatia},\ and\ \citenamefont {Tseng}}]{Keung:2020uew}%
  \BibitemOpen
  \bibfield  {author} {\bibinfo {author} {\bibfnamefont {W.-Y.}\ \bibnamefont
  {Keung}}, \bibinfo {author} {\bibfnamefont {D.}~\bibnamefont {Marfatia}}, \
  and\ \bibinfo {author} {\bibfnamefont {P.-Y.}\ \bibnamefont {Tseng}},\ }\href
  {\doibase 10.1016/j.jheap.2021.02.001} {\bibfield  {journal} {\bibinfo
  {journal} {JHEAp}\ }\textbf {\bibinfo {volume} {30}},\ \bibinfo {pages} {9}
  (\bibinfo {year} {2021})},\ \Eprint {http://arxiv.org/abs/2009.04444}
  {arXiv:2009.04444 [hep-ph]} \BibitemShut {NoStop}%
\bibitem [{\citenamefont {Aboubrahim}\ \emph {et~al.}(2021)\citenamefont
  {Aboubrahim}, \citenamefont {Klasen},\ and\ \citenamefont
  {Nath}}]{Aboubrahim:2020iwb}%
  \BibitemOpen
  \bibfield  {author} {\bibinfo {author} {\bibfnamefont {A.}~\bibnamefont
  {Aboubrahim}}, \bibinfo {author} {\bibfnamefont {M.}~\bibnamefont {Klasen}},
  \ and\ \bibinfo {author} {\bibfnamefont {P.}~\bibnamefont {Nath}},\ }\href
  {\doibase 10.1007/JHEP02(2021)229} {\bibfield  {journal} {\bibinfo  {journal}
  {JHEP}\ }\textbf {\bibinfo {volume} {02}},\ \bibinfo {pages} {229} (\bibinfo
  {year} {2021})},\ \Eprint {http://arxiv.org/abs/2011.08053} {arXiv:2011.08053
  [hep-ph]} \BibitemShut {NoStop}%
\bibitem [{\citenamefont {Dror}\ \emph
  {et~al.}(2020{\natexlab{a}})\citenamefont {Dror}, \citenamefont {Elor},\ and\
  \citenamefont {Mcgehee}}]{Dror:2019onn}%
  \BibitemOpen
  \bibfield  {author} {\bibinfo {author} {\bibfnamefont {J.~A.}\ \bibnamefont
  {Dror}}, \bibinfo {author} {\bibfnamefont {G.}~\bibnamefont {Elor}}, \ and\
  \bibinfo {author} {\bibfnamefont {R.}~\bibnamefont {Mcgehee}},\ }\href
  {\doibase 10.1103/PhysRevLett.124.181301} {\bibfield  {journal} {\bibinfo
  {journal} {Phys. Rev. Lett.}\ }\textbf {\bibinfo {volume} {124}},\ \bibinfo
  {pages} {18} (\bibinfo {year} {2020}{\natexlab{a}})},\ \Eprint
  {http://arxiv.org/abs/1905.12635} {arXiv:1905.12635 [hep-ph]} \BibitemShut
  {NoStop}%
\bibitem [{\citenamefont {Dror}\ \emph
  {et~al.}(2020{\natexlab{b}})\citenamefont {Dror}, \citenamefont {Elor},\ and\
  \citenamefont {Mcgehee}}]{Dror:2019dib}%
  \BibitemOpen
  \bibfield  {author} {\bibinfo {author} {\bibfnamefont {J.~A.}\ \bibnamefont
  {Dror}}, \bibinfo {author} {\bibfnamefont {G.}~\bibnamefont {Elor}}, \ and\
  \bibinfo {author} {\bibfnamefont {R.}~\bibnamefont {Mcgehee}},\ }\href
  {\doibase 10.1007/JHEP02(2020)134} {\bibfield  {journal} {\bibinfo  {journal}
  {JHEP}\ }\textbf {\bibinfo {volume} {02}},\ \bibinfo {pages} {134} (\bibinfo
  {year} {2020}{\natexlab{b}})},\ \Eprint {http://arxiv.org/abs/1908.10861}
  {arXiv:1908.10861 [hep-ph]} \BibitemShut {NoStop}%
\bibitem [{\citenamefont {Dror}\ \emph {et~al.}(2021)\citenamefont {Dror},
  \citenamefont {Elor}, \citenamefont {McGehee},\ and\ \citenamefont
  {Yu}}]{Dror:2020czw}%
  \BibitemOpen
  \bibfield  {author} {\bibinfo {author} {\bibfnamefont {J.~A.}\ \bibnamefont
  {Dror}}, \bibinfo {author} {\bibfnamefont {G.}~\bibnamefont {Elor}}, \bibinfo
  {author} {\bibfnamefont {R.}~\bibnamefont {McGehee}}, \ and\ \bibinfo
  {author} {\bibfnamefont {T.-T.}\ \bibnamefont {Yu}},\ }\href {\doibase
  10.1103/PhysRevD.103.035001} {\bibfield  {journal} {\bibinfo  {journal}
  {Phys. Rev. D}\ }\textbf {\bibinfo {volume} {103}},\ \bibinfo {pages}
  {035001} (\bibinfo {year} {2021})},\ \Eprint
  {http://arxiv.org/abs/2011.01940} {arXiv:2011.01940 [hep-ph]} \BibitemShut
  {NoStop}%
\bibitem [{\citenamefont {Kahn}\ and\ \citenamefont
  {Lin}(2021)}]{Kahn:2021ttr}%
  \BibitemOpen
  \bibfield  {author} {\bibinfo {author} {\bibfnamefont {Y.}~\bibnamefont
  {Kahn}}\ and\ \bibinfo {author} {\bibfnamefont {T.}~\bibnamefont {Lin}},\
  }\href@noop {} {\  (\bibinfo {year} {2021})},\ \Eprint
  {http://arxiv.org/abs/2108.03239} {arXiv:2108.03239 [hep-ph]} \BibitemShut
  {NoStop}%
\bibitem [{\citenamefont {Liang}\ \emph {et~al.}(2021)\citenamefont {Liang},
  \citenamefont {Mo},\ and\ \citenamefont {Zhang}}]{Liang:2021zkg}%
  \BibitemOpen
  \bibfield  {author} {\bibinfo {author} {\bibfnamefont {Z.-L.}\ \bibnamefont
  {Liang}}, \bibinfo {author} {\bibfnamefont {C.}~\bibnamefont {Mo}}, \ and\
  \bibinfo {author} {\bibfnamefont {P.}~\bibnamefont {Zhang}},\ }\href@noop {}
  {\  (\bibinfo {year} {2021})},\ \Eprint {http://arxiv.org/abs/2107.01209}
  {arXiv:2107.01209 [hep-ph]} \BibitemShut {NoStop}%
\bibitem [{\citenamefont {Andersson}\ \emph {et~al.}(2020)\citenamefont
  {Andersson}, \citenamefont {B\"okmark}, \citenamefont {Catena}, \citenamefont
  {Emken}, \citenamefont {Moberg},\ and\ \citenamefont
  {\r{A}strand}}]{Andersson:2020uwc}%
  \BibitemOpen
  \bibfield  {author} {\bibinfo {author} {\bibfnamefont {E.}~\bibnamefont
  {Andersson}}, \bibinfo {author} {\bibfnamefont {A.}~\bibnamefont
  {B\"okmark}}, \bibinfo {author} {\bibfnamefont {R.}~\bibnamefont {Catena}},
  \bibinfo {author} {\bibfnamefont {T.}~\bibnamefont {Emken}}, \bibinfo
  {author} {\bibfnamefont {H.~K.}\ \bibnamefont {Moberg}}, \ and\ \bibinfo
  {author} {\bibfnamefont {E.}~\bibnamefont {\r{A}strand}},\ }\href {\doibase
  10.1088/1475-7516/2020/05/036} {\bibfield  {journal} {\bibinfo  {journal}
  {JCAP}\ }\textbf {\bibinfo {volume} {05}},\ \bibinfo {pages} {036} (\bibinfo
  {year} {2020})},\ \Eprint {http://arxiv.org/abs/2001.08910} {arXiv:2001.08910
  [hep-ph]} \BibitemShut {NoStop}%
\bibitem [{\citenamefont {Graham}\ \emph {et~al.}(2012)\citenamefont {Graham},
  \citenamefont {Kaplan}, \citenamefont {Rajendran},\ and\ \citenamefont
  {Walters}}]{Graham:2012su}%
  \BibitemOpen
  \bibfield  {author} {\bibinfo {author} {\bibfnamefont {P.~W.}\ \bibnamefont
  {Graham}}, \bibinfo {author} {\bibfnamefont {D.~E.}\ \bibnamefont {Kaplan}},
  \bibinfo {author} {\bibfnamefont {S.}~\bibnamefont {Rajendran}}, \ and\
  \bibinfo {author} {\bibfnamefont {M.~T.}\ \bibnamefont {Walters}},\ }\href
  {\doibase 10.1016/j.dark.2012.09.001} {\bibfield  {journal} {\bibinfo
  {journal} {Phys. Dark Univ.}\ }\textbf {\bibinfo {volume} {1}},\ \bibinfo
  {pages} {32} (\bibinfo {year} {2012})},\ \Eprint
  {http://arxiv.org/abs/1203.2531} {arXiv:1203.2531 [hep-ph]} \BibitemShut
  {NoStop}%
\bibitem [{\citenamefont {Essig}\ \emph {et~al.}(2016)\citenamefont {Essig},
  \citenamefont {Fernandez-Serra}, \citenamefont {Mardon}, \citenamefont
  {Soto}, \citenamefont {Volansky},\ and\ \citenamefont {Yu}}]{Essig:2015cda}%
  \BibitemOpen
  \bibfield  {author} {\bibinfo {author} {\bibfnamefont {R.}~\bibnamefont
  {Essig}}, \bibinfo {author} {\bibfnamefont {M.}~\bibnamefont
  {Fernandez-Serra}}, \bibinfo {author} {\bibfnamefont {J.}~\bibnamefont
  {Mardon}}, \bibinfo {author} {\bibfnamefont {A.}~\bibnamefont {Soto}},
  \bibinfo {author} {\bibfnamefont {T.}~\bibnamefont {Volansky}}, \ and\
  \bibinfo {author} {\bibfnamefont {T.-T.}\ \bibnamefont {Yu}},\ }\href
  {\doibase 10.1007/JHEP05(2016)046} {\bibfield  {journal} {\bibinfo  {journal}
  {JHEP}\ }\textbf {\bibinfo {volume} {05}},\ \bibinfo {pages} {046} (\bibinfo
  {year} {2016})},\ \Eprint {http://arxiv.org/abs/1509.01598} {arXiv:1509.01598
  [hep-ph]} \BibitemShut {NoStop}%
\bibitem [{\citenamefont {Lazanu}\ \emph {et~al.}(2013)\citenamefont {Lazanu},
  \citenamefont {Ciurea},\ and\ \citenamefont {Lazanu}}]{Lazanu:2012fi}%
  \BibitemOpen
  \bibfield  {author} {\bibinfo {author} {\bibfnamefont {I.}~\bibnamefont
  {Lazanu}}, \bibinfo {author} {\bibfnamefont {M.~L.}\ \bibnamefont {Ciurea}},
  \ and\ \bibinfo {author} {\bibfnamefont {S.}~\bibnamefont {Lazanu}},\ }\href
  {\doibase 10.1016/j.astropartphys.2013.01.005} {\bibfield  {journal}
  {\bibinfo  {journal} {Astropart. Phys.}\ }\textbf {\bibinfo {volume} {44}},\
  \bibinfo {pages} {9} (\bibinfo {year} {2013})},\ \Eprint
  {http://arxiv.org/abs/1211.1369} {arXiv:1211.1369 [astro-ph.IM]} \BibitemShut
  {NoStop}%
\bibitem [{\citenamefont {Lehnert}\ \emph {et~al.}(2020)\citenamefont
  {Lehnert}, \citenamefont {Ramani}, \citenamefont {Hult}, \citenamefont
  {Lutter}, \citenamefont {Pospelov}, \citenamefont {Rajendran},\ and\
  \citenamefont {Zuber}}]{Lehnert:2019tuw}%
  \BibitemOpen
  \bibfield  {author} {\bibinfo {author} {\bibfnamefont {B.}~\bibnamefont
  {Lehnert}}, \bibinfo {author} {\bibfnamefont {H.}~\bibnamefont {Ramani}},
  \bibinfo {author} {\bibfnamefont {M.}~\bibnamefont {Hult}}, \bibinfo {author}
  {\bibfnamefont {G.}~\bibnamefont {Lutter}}, \bibinfo {author} {\bibfnamefont
  {M.}~\bibnamefont {Pospelov}}, \bibinfo {author} {\bibfnamefont
  {S.}~\bibnamefont {Rajendran}}, \ and\ \bibinfo {author} {\bibfnamefont
  {K.}~\bibnamefont {Zuber}},\ }\href {\doibase 10.1103/PhysRevLett.124.181802}
  {\bibfield  {journal} {\bibinfo  {journal} {Phys. Rev. Lett.}\ }\textbf
  {\bibinfo {volume} {124}},\ \bibinfo {pages} {181802} (\bibinfo {year}
  {2020})},\ \Eprint {http://arxiv.org/abs/1911.07865} {arXiv:1911.07865
  [astro-ph.CO]} \BibitemShut {NoStop}%
\bibitem [{\citenamefont {Essig}\ \emph {et~al.}(2020)\citenamefont {Essig},
  \citenamefont {Pradler}, \citenamefont {Sholapurkar},\ and\ \citenamefont
  {Yu}}]{Essig:2019xkx}%
  \BibitemOpen
  \bibfield  {author} {\bibinfo {author} {\bibfnamefont {R.}~\bibnamefont
  {Essig}}, \bibinfo {author} {\bibfnamefont {J.}~\bibnamefont {Pradler}},
  \bibinfo {author} {\bibfnamefont {M.}~\bibnamefont {Sholapurkar}}, \ and\
  \bibinfo {author} {\bibfnamefont {T.-T.}\ \bibnamefont {Yu}},\ }\href
  {\doibase 10.1103/PhysRevLett.124.021801} {\bibfield  {journal} {\bibinfo
  {journal} {Phys. Rev. Lett.}\ }\textbf {\bibinfo {volume} {124}},\ \bibinfo
  {pages} {021801} (\bibinfo {year} {2020})},\ \Eprint
  {http://arxiv.org/abs/1908.10881} {arXiv:1908.10881 [hep-ph]} \BibitemShut
  {NoStop}%
\bibitem [{\citenamefont {Knapen}\ \emph {et~al.}(2021)\citenamefont {Knapen},
  \citenamefont {Kozaczuk},\ and\ \citenamefont {Lin}}]{Knapen:2020aky}%
  \BibitemOpen
  \bibfield  {author} {\bibinfo {author} {\bibfnamefont {S.}~\bibnamefont
  {Knapen}}, \bibinfo {author} {\bibfnamefont {J.}~\bibnamefont {Kozaczuk}}, \
  and\ \bibinfo {author} {\bibfnamefont {T.}~\bibnamefont {Lin}},\ }\href
  {\doibase 10.1103/PhysRevLett.127.081805} {\bibfield  {journal} {\bibinfo
  {journal} {Phys. Rev. Lett.}\ }\textbf {\bibinfo {volume} {127}},\ \bibinfo
  {pages} {081805} (\bibinfo {year} {2021})},\ \Eprint
  {http://arxiv.org/abs/2011.09496} {arXiv:2011.09496 [hep-ph]} \BibitemShut
  {NoStop}%
\bibitem [{\citenamefont {Ibe}\ \emph {et~al.}(2018)\citenamefont {Ibe},
  \citenamefont {Nakano}, \citenamefont {Shoji},\ and\ \citenamefont
  {Suzuki}}]{Ibe:2017yqa}%
  \BibitemOpen
  \bibfield  {author} {\bibinfo {author} {\bibfnamefont {M.}~\bibnamefont
  {Ibe}}, \bibinfo {author} {\bibfnamefont {W.}~\bibnamefont {Nakano}},
  \bibinfo {author} {\bibfnamefont {Y.}~\bibnamefont {Shoji}}, \ and\ \bibinfo
  {author} {\bibfnamefont {K.}~\bibnamefont {Suzuki}},\ }\href {\doibase
  10.1007/JHEP03(2018)194} {\bibfield  {journal} {\bibinfo  {journal} {JHEP}\
  }\textbf {\bibinfo {volume} {03}},\ \bibinfo {pages} {194} (\bibinfo {year}
  {2018})},\ \Eprint {http://arxiv.org/abs/1707.07258} {arXiv:1707.07258
  [hep-ph]} \BibitemShut {NoStop}%
\bibitem [{\citenamefont {Baxter}\ \emph {et~al.}(2020)\citenamefont {Baxter},
  \citenamefont {Kahn},\ and\ \citenamefont {Krnjaic}}]{Baxter:2019pnz}%
  \BibitemOpen
  \bibfield  {author} {\bibinfo {author} {\bibfnamefont {D.}~\bibnamefont
  {Baxter}}, \bibinfo {author} {\bibfnamefont {Y.}~\bibnamefont {Kahn}}, \ and\
  \bibinfo {author} {\bibfnamefont {G.}~\bibnamefont {Krnjaic}},\ }\href
  {\doibase 10.1103/PhysRevD.101.076014} {\bibfield  {journal} {\bibinfo
  {journal} {Phys. Rev. D}\ }\textbf {\bibinfo {volume} {101}},\ \bibinfo
  {pages} {076014} (\bibinfo {year} {2020})},\ \Eprint
  {http://arxiv.org/abs/1908.00012} {arXiv:1908.00012 [hep-ph]} \BibitemShut
  {NoStop}%
\bibitem [{\citenamefont {Bell}\ \emph {et~al.}(2021)\citenamefont {Bell},
  \citenamefont {Dent}, \citenamefont {Dutta}, \citenamefont {Ghosh},
  \citenamefont {Kumar},\ and\ \citenamefont {Newstead}}]{Bell:2021zkr}%
  \BibitemOpen
  \bibfield  {author} {\bibinfo {author} {\bibfnamefont {N.~F.}\ \bibnamefont
  {Bell}}, \bibinfo {author} {\bibfnamefont {J.~B.}\ \bibnamefont {Dent}},
  \bibinfo {author} {\bibfnamefont {B.}~\bibnamefont {Dutta}}, \bibinfo
  {author} {\bibfnamefont {S.}~\bibnamefont {Ghosh}}, \bibinfo {author}
  {\bibfnamefont {J.}~\bibnamefont {Kumar}}, \ and\ \bibinfo {author}
  {\bibfnamefont {J.~L.}\ \bibnamefont {Newstead}},\ }\href@noop {} {\
  (\bibinfo {year} {2021})},\ \Eprint {http://arxiv.org/abs/2103.05890}
  {arXiv:2103.05890 [hep-ph]} \BibitemShut {NoStop}%
\bibitem [{\citenamefont {Hochberg}\ \emph {et~al.}(2021)\citenamefont
  {Hochberg}, \citenamefont {Kramer}, \citenamefont {Kurinsky},\ and\
  \citenamefont {Lehmann}}]{Hochberg:2021ymx}%
  \BibitemOpen
  \bibfield  {author} {\bibinfo {author} {\bibfnamefont {Y.}~\bibnamefont
  {Hochberg}}, \bibinfo {author} {\bibfnamefont {E.~D.}\ \bibnamefont
  {Kramer}}, \bibinfo {author} {\bibfnamefont {N.}~\bibnamefont {Kurinsky}}, \
  and\ \bibinfo {author} {\bibfnamefont {B.~V.}\ \bibnamefont {Lehmann}},\
  }\href@noop {} {\  (\bibinfo {year} {2021})},\ \Eprint
  {http://arxiv.org/abs/2109.04473} {arXiv:2109.04473 [hep-ph]} \BibitemShut
  {NoStop}%
\bibitem [{\citenamefont {Smirnov}\ and\ \citenamefont
  {Beacom}(2020)}]{Smirnov:2020zwf}%
  \BibitemOpen
  \bibfield  {author} {\bibinfo {author} {\bibfnamefont {J.}~\bibnamefont
  {Smirnov}}\ and\ \bibinfo {author} {\bibfnamefont {J.~F.}\ \bibnamefont
  {Beacom}},\ }\href {\doibase 10.1103/PhysRevLett.125.131301} {\bibfield
  {journal} {\bibinfo  {journal} {Phys. Rev. Lett.}\ }\textbf {\bibinfo
  {volume} {125}},\ \bibinfo {pages} {131301} (\bibinfo {year} {2020})},\
  \Eprint {http://arxiv.org/abs/2002.04038} {arXiv:2002.04038 [hep-ph]}
  \BibitemShut {NoStop}%
\bibitem [{\citenamefont {Kuflik}\ \emph {et~al.}(2016)\citenamefont {Kuflik},
  \citenamefont {Perelstein}, \citenamefont {Lorier},\ and\ \citenamefont
  {Tsai}}]{Kuflik:2015isi}%
  \BibitemOpen
  \bibfield  {author} {\bibinfo {author} {\bibfnamefont {E.}~\bibnamefont
  {Kuflik}}, \bibinfo {author} {\bibfnamefont {M.}~\bibnamefont {Perelstein}},
  \bibinfo {author} {\bibfnamefont {N.~R.-L.}\ \bibnamefont {Lorier}}, \ and\
  \bibinfo {author} {\bibfnamefont {Y.-D.}\ \bibnamefont {Tsai}},\ }\href
  {\doibase 10.1103/PhysRevLett.116.221302} {\bibfield  {journal} {\bibinfo
  {journal} {Phys. Rev. Lett.}\ }\textbf {\bibinfo {volume} {116}},\ \bibinfo
  {pages} {221302} (\bibinfo {year} {2016})},\ \Eprint
  {http://arxiv.org/abs/1512.04545} {arXiv:1512.04545 [hep-ph]} \BibitemShut
  {NoStop}%
\bibitem [{\citenamefont {Hochberg}\ \emph {et~al.}(2014)\citenamefont
  {Hochberg}, \citenamefont {Kuflik}, \citenamefont {Volansky},\ and\
  \citenamefont {Wacker}}]{Hochberg:2014dra}%
  \BibitemOpen
  \bibfield  {author} {\bibinfo {author} {\bibfnamefont {Y.}~\bibnamefont
  {Hochberg}}, \bibinfo {author} {\bibfnamefont {E.}~\bibnamefont {Kuflik}},
  \bibinfo {author} {\bibfnamefont {T.}~\bibnamefont {Volansky}}, \ and\
  \bibinfo {author} {\bibfnamefont {J.~G.}\ \bibnamefont {Wacker}},\ }\href
  {\doibase 10.1103/PhysRevLett.113.171301} {\bibfield  {journal} {\bibinfo
  {journal} {Phys. Rev. Lett.}\ }\textbf {\bibinfo {volume} {113}},\ \bibinfo
  {pages} {171301} (\bibinfo {year} {2014})},\ \Eprint
  {http://arxiv.org/abs/1402.5143} {arXiv:1402.5143 [hep-ph]} \BibitemShut
  {NoStop}%
\bibitem [{\citenamefont {Hochberg}\ \emph {et~al.}(2015)\citenamefont
  {Hochberg}, \citenamefont {Kuflik}, \citenamefont {Murayama}, \citenamefont
  {Volansky},\ and\ \citenamefont {Wacker}}]{Hochberg:2014kqa}%
  \BibitemOpen
  \bibfield  {author} {\bibinfo {author} {\bibfnamefont {Y.}~\bibnamefont
  {Hochberg}}, \bibinfo {author} {\bibfnamefont {E.}~\bibnamefont {Kuflik}},
  \bibinfo {author} {\bibfnamefont {H.}~\bibnamefont {Murayama}}, \bibinfo
  {author} {\bibfnamefont {T.}~\bibnamefont {Volansky}}, \ and\ \bibinfo
  {author} {\bibfnamefont {J.~G.}\ \bibnamefont {Wacker}},\ }\href {\doibase
  10.1103/PhysRevLett.115.021301} {\bibfield  {journal} {\bibinfo  {journal}
  {Phys. Rev. Lett.}\ }\textbf {\bibinfo {volume} {115}},\ \bibinfo {pages}
  {021301} (\bibinfo {year} {2015})},\ \Eprint {http://arxiv.org/abs/1411.3727}
  {arXiv:1411.3727 [hep-ph]} \BibitemShut {NoStop}%
\bibitem [{\citenamefont {Cline}\ \emph {et~al.}(2017)\citenamefont {Cline},
  \citenamefont {Liu}, \citenamefont {Slatyer},\ and\ \citenamefont
  {Xue}}]{Cline:2017tka}%
  \BibitemOpen
  \bibfield  {author} {\bibinfo {author} {\bibfnamefont {J.~M.}\ \bibnamefont
  {Cline}}, \bibinfo {author} {\bibfnamefont {H.}~\bibnamefont {Liu}}, \bibinfo
  {author} {\bibfnamefont {T.}~\bibnamefont {Slatyer}}, \ and\ \bibinfo
  {author} {\bibfnamefont {W.}~\bibnamefont {Xue}},\ }\href {\doibase
  10.1103/PhysRevD.96.083521} {\bibfield  {journal} {\bibinfo  {journal} {Phys.
  Rev. D}\ }\textbf {\bibinfo {volume} {96}},\ \bibinfo {pages} {083521}
  (\bibinfo {year} {2017})},\ \Eprint {http://arxiv.org/abs/1702.07716}
  {arXiv:1702.07716 [hep-ph]} \BibitemShut {NoStop}%
\bibitem [{\citenamefont {Akerib}\ \emph {et~al.}(2017)\citenamefont {Akerib}
  \emph {et~al.}}]{LUX:2016ggv}%
  \BibitemOpen
  \bibfield  {author} {\bibinfo {author} {\bibfnamefont {D.~S.}\ \bibnamefont
  {Akerib}} \emph {et~al.} (\bibinfo {collaboration} {LUX}),\ }\href {\doibase
  10.1103/PhysRevLett.118.021303} {\bibfield  {journal} {\bibinfo  {journal}
  {Phys. Rev. Lett.}\ }\textbf {\bibinfo {volume} {118}},\ \bibinfo {pages}
  {021303} (\bibinfo {year} {2017})},\ \Eprint
  {http://arxiv.org/abs/1608.07648} {arXiv:1608.07648 [astro-ph.CO]}
  \BibitemShut {NoStop}%
\bibitem [{\citenamefont {Cui}\ \emph {et~al.}(2017{\natexlab{b}})\citenamefont
  {Cui} \emph {et~al.}}]{PandaX-II:2017hlx}%
  \BibitemOpen
  \bibfield  {author} {\bibinfo {author} {\bibfnamefont {X.}~\bibnamefont
  {Cui}} \emph {et~al.} (\bibinfo {collaboration} {PandaX-II}),\ }\href
  {\doibase 10.1103/PhysRevLett.119.181302} {\bibfield  {journal} {\bibinfo
  {journal} {Phys. Rev. Lett.}\ }\textbf {\bibinfo {volume} {119}},\ \bibinfo
  {pages} {181302} (\bibinfo {year} {2017}{\natexlab{b}})},\ \Eprint
  {http://arxiv.org/abs/1708.06917} {arXiv:1708.06917 [astro-ph.CO]}
  \BibitemShut {NoStop}%
\bibitem [{\citenamefont {Aprile}\ \emph
  {et~al.}(2018{\natexlab{b}})\citenamefont {Aprile} \emph
  {et~al.}}]{XENON:2018voc}%
  \BibitemOpen
  \bibfield  {author} {\bibinfo {author} {\bibfnamefont {E.}~\bibnamefont
  {Aprile}} \emph {et~al.} (\bibinfo {collaboration} {XENON}),\ }\href
  {\doibase 10.1103/PhysRevLett.121.111302} {\bibfield  {journal} {\bibinfo
  {journal} {Phys. Rev. Lett.}\ }\textbf {\bibinfo {volume} {121}},\ \bibinfo
  {pages} {111302} (\bibinfo {year} {2018}{\natexlab{b}})},\ \Eprint
  {http://arxiv.org/abs/1805.12562} {arXiv:1805.12562 [astro-ph.CO]}
  \BibitemShut {NoStop}%
\bibitem [{\citenamefont {Navarro}\ \emph {et~al.}(1997)\citenamefont
  {Navarro}, \citenamefont {Frenk},\ and\ \citenamefont
  {White}}]{Navarro:1996gj}%
  \BibitemOpen
  \bibfield  {author} {\bibinfo {author} {\bibfnamefont {J.~F.}\ \bibnamefont
  {Navarro}}, \bibinfo {author} {\bibfnamefont {C.~S.}\ \bibnamefont {Frenk}},
  \ and\ \bibinfo {author} {\bibfnamefont {S.~D.~M.}\ \bibnamefont {White}},\
  }\href {\doibase 10.1086/304888} {\bibfield  {journal} {\bibinfo  {journal}
  {Astrophys. J.}\ }\textbf {\bibinfo {volume} {490}},\ \bibinfo {pages} {493}
  (\bibinfo {year} {1997})},\ \Eprint {http://arxiv.org/abs/astro-ph/9611107}
  {arXiv:astro-ph/9611107} \BibitemShut {NoStop}%
\bibitem [{\citenamefont {Cao}\ \emph {et~al.}(2021)\citenamefont {Cao},
  \citenamefont {Ding},\ and\ \citenamefont {Xiang}}]{Cao:2020bwd}%
  \BibitemOpen
  \bibfield  {author} {\bibinfo {author} {\bibfnamefont {Q.-H.}\ \bibnamefont
  {Cao}}, \bibinfo {author} {\bibfnamefont {R.}~\bibnamefont {Ding}}, \ and\
  \bibinfo {author} {\bibfnamefont {Q.-F.}\ \bibnamefont {Xiang}},\ }\href
  {\doibase 10.1088/1674-1137/abe195} {\bibfield  {journal} {\bibinfo
  {journal} {Chin. Phys. C}\ }\textbf {\bibinfo {volume} {45}},\ \bibinfo
  {pages} {045002} (\bibinfo {year} {2021})},\ \Eprint
  {http://arxiv.org/abs/2006.12767} {arXiv:2006.12767 [hep-ph]} \BibitemShut
  {NoStop}%
\bibitem [{\citenamefont {Kopp}\ \emph {et~al.}(2009)\citenamefont {Kopp},
  \citenamefont {Niro}, \citenamefont {Schwetz},\ and\ \citenamefont
  {Zupan}}]{Kopp:2009et}%
  \BibitemOpen
  \bibfield  {author} {\bibinfo {author} {\bibfnamefont {J.}~\bibnamefont
  {Kopp}}, \bibinfo {author} {\bibfnamefont {V.}~\bibnamefont {Niro}}, \bibinfo
  {author} {\bibfnamefont {T.}~\bibnamefont {Schwetz}}, \ and\ \bibinfo
  {author} {\bibfnamefont {J.}~\bibnamefont {Zupan}},\ }\href {\doibase
  10.1103/PhysRevD.80.083502} {\bibfield  {journal} {\bibinfo  {journal} {Phys.
  Rev. D}\ }\textbf {\bibinfo {volume} {80}},\ \bibinfo {pages} {083502}
  (\bibinfo {year} {2009})},\ \Eprint {http://arxiv.org/abs/0907.3159}
  {arXiv:0907.3159 [hep-ph]} \BibitemShut {NoStop}%
\bibitem [{\citenamefont {Bunge}\ \emph {et~al.}(1993)\citenamefont {Bunge},
  \citenamefont {Barrientos},\ and\ \citenamefont {Bunge}}]{Bunge:1993jsz}%
  \BibitemOpen
  \bibfield  {author} {\bibinfo {author} {\bibfnamefont {C.~F.}\ \bibnamefont
  {Bunge}}, \bibinfo {author} {\bibfnamefont {J.~A.}\ \bibnamefont
  {Barrientos}}, \ and\ \bibinfo {author} {\bibfnamefont {A.~V.}\ \bibnamefont
  {Bunge}},\ }\href {\doibase 10.1006/adnd.1993.1003} {\bibfield  {journal}
  {\bibinfo  {journal} {Atom. Data Nucl. Data Tabl.}\ }\textbf {\bibinfo
  {volume} {53}},\ \bibinfo {pages} {113} (\bibinfo {year} {1993})}\BibitemShut
  {NoStop}%
\bibitem [{\citenamefont {Aprile}\ \emph {et~al.}(2020)\citenamefont {Aprile}
  \emph {et~al.}}]{Aprile:2020tmw}%
  \BibitemOpen
  \bibfield  {author} {\bibinfo {author} {\bibfnamefont {E.}~\bibnamefont
  {Aprile}} \emph {et~al.} (\bibinfo {collaboration} {XENON}),\ }\href
  {\doibase 10.1103/PhysRevD.102.072004} {\bibfield  {journal} {\bibinfo
  {journal} {Phys. Rev. D}\ }\textbf {\bibinfo {volume} {102}},\ \bibinfo
  {pages} {072004} (\bibinfo {year} {2020})},\ \Eprint
  {http://arxiv.org/abs/2006.09721} {arXiv:2006.09721 [hep-ex]} \BibitemShut
  {NoStop}%
\bibitem [{\citenamefont {Cirelli}\ \emph {et~al.}(2006)\citenamefont
  {Cirelli}, \citenamefont {Fornengo},\ and\ \citenamefont
  {Strumia}}]{Cirelli:2005uq}%
  \BibitemOpen
  \bibfield  {author} {\bibinfo {author} {\bibfnamefont {M.}~\bibnamefont
  {Cirelli}}, \bibinfo {author} {\bibfnamefont {N.}~\bibnamefont {Fornengo}}, \
  and\ \bibinfo {author} {\bibfnamefont {A.}~\bibnamefont {Strumia}},\ }\href
  {\doibase 10.1016/j.nuclphysb.2006.07.012} {\bibfield  {journal} {\bibinfo
  {journal} {Nucl. Phys. B}\ }\textbf {\bibinfo {volume} {753}},\ \bibinfo
  {pages} {178} (\bibinfo {year} {2006})},\ \Eprint
  {http://arxiv.org/abs/hep-ph/0512090} {arXiv:hep-ph/0512090} \BibitemShut
  {NoStop}%
\bibitem [{\citenamefont {Lopez~Honorez}\ \emph {et~al.}(2007)\citenamefont
  {Lopez~Honorez}, \citenamefont {Nezri}, \citenamefont {Oliver},\ and\
  \citenamefont {Tytgat}}]{LopezHonorez:2006gr}%
  \BibitemOpen
  \bibfield  {author} {\bibinfo {author} {\bibfnamefont {L.}~\bibnamefont
  {Lopez~Honorez}}, \bibinfo {author} {\bibfnamefont {E.}~\bibnamefont
  {Nezri}}, \bibinfo {author} {\bibfnamefont {J.~F.}\ \bibnamefont {Oliver}}, \
  and\ \bibinfo {author} {\bibfnamefont {M.~H.~G.}\ \bibnamefont {Tytgat}},\
  }\href {\doibase 10.1088/1475-7516/2007/02/028} {\bibfield  {journal}
  {\bibinfo  {journal} {JCAP}\ }\textbf {\bibinfo {volume} {02}},\ \bibinfo
  {pages} {028} (\bibinfo {year} {2007})},\ \Eprint
  {http://arxiv.org/abs/hep-ph/0612275} {arXiv:hep-ph/0612275} \BibitemShut
  {NoStop}%
\bibitem [{\citenamefont {Petraki}\ \emph {et~al.}(2015)\citenamefont
  {Petraki}, \citenamefont {Postma},\ and\ \citenamefont
  {Wiechers}}]{Petraki:2015hla}%
  \BibitemOpen
  \bibfield  {author} {\bibinfo {author} {\bibfnamefont {K.}~\bibnamefont
  {Petraki}}, \bibinfo {author} {\bibfnamefont {M.}~\bibnamefont {Postma}}, \
  and\ \bibinfo {author} {\bibfnamefont {M.}~\bibnamefont {Wiechers}},\ }\href
  {\doibase 10.1007/JHEP06(2015)128} {\bibfield  {journal} {\bibinfo  {journal}
  {JHEP}\ }\textbf {\bibinfo {volume} {06}},\ \bibinfo {pages} {128} (\bibinfo
  {year} {2015})},\ \Eprint {http://arxiv.org/abs/1505.00109} {arXiv:1505.00109
  [hep-ph]} \BibitemShut {NoStop}%
\bibitem [{\citenamefont {Petraki}\ \emph {et~al.}(2017)\citenamefont
  {Petraki}, \citenamefont {Postma},\ and\ \citenamefont
  {de~Vries}}]{Petraki:2016cnz}%
  \BibitemOpen
  \bibfield  {author} {\bibinfo {author} {\bibfnamefont {K.}~\bibnamefont
  {Petraki}}, \bibinfo {author} {\bibfnamefont {M.}~\bibnamefont {Postma}}, \
  and\ \bibinfo {author} {\bibfnamefont {J.}~\bibnamefont {de~Vries}},\ }\href
  {\doibase 10.1007/JHEP04(2017)077} {\bibfield  {journal} {\bibinfo  {journal}
  {JHEP}\ }\textbf {\bibinfo {volume} {04}},\ \bibinfo {pages} {077} (\bibinfo
  {year} {2017})},\ \Eprint {http://arxiv.org/abs/1611.01394} {arXiv:1611.01394
  [hep-ph]} \BibitemShut {NoStop}%
\bibitem [{\citenamefont {Caputo}\ \emph {et~al.}(2021)\citenamefont {Caputo},
  \citenamefont {Millar}, \citenamefont {O'Hare},\ and\ \citenamefont
  {Vitagliano}}]{Caputo:2021eaa}%
  \BibitemOpen
  \bibfield  {author} {\bibinfo {author} {\bibfnamefont {A.}~\bibnamefont
  {Caputo}}, \bibinfo {author} {\bibfnamefont {A.~J.}\ \bibnamefont {Millar}},
  \bibinfo {author} {\bibfnamefont {C.~A.~J.}\ \bibnamefont {O'Hare}}, \ and\
  \bibinfo {author} {\bibfnamefont {E.}~\bibnamefont {Vitagliano}},\
  }\href@noop {} {\  (\bibinfo {year} {2021})},\ \Eprint
  {http://arxiv.org/abs/2105.04565} {arXiv:2105.04565 [hep-ph]} \BibitemShut
  {NoStop}%
\bibitem [{\citenamefont {Giudice}\ \emph {et~al.}(2018)\citenamefont
  {Giudice}, \citenamefont {Kim}, \citenamefont {Park},\ and\ \citenamefont
  {Shin}}]{Giudice:2017zke}%
  \BibitemOpen
  \bibfield  {author} {\bibinfo {author} {\bibfnamefont {G.~F.}\ \bibnamefont
  {Giudice}}, \bibinfo {author} {\bibfnamefont {D.}~\bibnamefont {Kim}},
  \bibinfo {author} {\bibfnamefont {J.-C.}\ \bibnamefont {Park}}, \ and\
  \bibinfo {author} {\bibfnamefont {S.}~\bibnamefont {Shin}},\ }\href {\doibase
  10.1016/j.physletb.2018.03.043} {\bibfield  {journal} {\bibinfo  {journal}
  {Phys. Lett. B}\ }\textbf {\bibinfo {volume} {780}},\ \bibinfo {pages} {543}
  (\bibinfo {year} {2018})},\ \Eprint {http://arxiv.org/abs/1712.07126}
  {arXiv:1712.07126 [hep-ph]} \BibitemShut {NoStop}%
\bibitem [{\citenamefont {Agashe}\ \emph {et~al.}(2014)\citenamefont {Agashe},
  \citenamefont {Cui}, \citenamefont {Necib},\ and\ \citenamefont
  {Thaler}}]{Agashe:2014yua}%
  \BibitemOpen
  \bibfield  {author} {\bibinfo {author} {\bibfnamefont {K.}~\bibnamefont
  {Agashe}}, \bibinfo {author} {\bibfnamefont {Y.}~\bibnamefont {Cui}},
  \bibinfo {author} {\bibfnamefont {L.}~\bibnamefont {Necib}}, \ and\ \bibinfo
  {author} {\bibfnamefont {J.}~\bibnamefont {Thaler}},\ }\href {\doibase
  10.1088/1475-7516/2014/10/062} {\bibfield  {journal} {\bibinfo  {journal}
  {JCAP}\ }\textbf {\bibinfo {volume} {10}},\ \bibinfo {pages} {062} (\bibinfo
  {year} {2014})},\ \Eprint {http://arxiv.org/abs/1405.7370} {arXiv:1405.7370
  [hep-ph]} \BibitemShut {NoStop}%
\bibitem [{\citenamefont {Lewin}\ and\ \citenamefont
  {Smith}(1996)}]{Lewin:1995rx}%
  \BibitemOpen
  \bibfield  {author} {\bibinfo {author} {\bibfnamefont {J.~D.}\ \bibnamefont
  {Lewin}}\ and\ \bibinfo {author} {\bibfnamefont {P.~F.}\ \bibnamefont
  {Smith}},\ }\href {\doibase 10.1016/S0927-6505(96)00047-3} {\bibfield
  {journal} {\bibinfo  {journal} {Astropart. Phys.}\ }\textbf {\bibinfo
  {volume} {6}},\ \bibinfo {pages} {87} (\bibinfo {year} {1996})}\BibitemShut
  {NoStop}%
\bibitem [{\citenamefont {Amole}\ \emph {et~al.}(2017)\citenamefont {Amole}
  \emph {et~al.}}]{PICO:2017tgi}%
  \BibitemOpen
  \bibfield  {author} {\bibinfo {author} {\bibfnamefont {C.}~\bibnamefont
  {Amole}} \emph {et~al.} (\bibinfo {collaboration} {PICO}),\ }\href {\doibase
  10.1103/PhysRevLett.118.251301} {\bibfield  {journal} {\bibinfo  {journal}
  {Phys. Rev. Lett.}\ }\textbf {\bibinfo {volume} {118}},\ \bibinfo {pages}
  {251301} (\bibinfo {year} {2017})},\ \Eprint
  {http://arxiv.org/abs/1702.07666} {arXiv:1702.07666 [astro-ph.CO]}
  \BibitemShut {NoStop}%
\bibitem [{\citenamefont {Agnese}\ \emph {et~al.}(2014)\citenamefont {Agnese}
  \emph {et~al.}}]{SuperCDMS:2014cds}%
  \BibitemOpen
  \bibfield  {author} {\bibinfo {author} {\bibfnamefont {R.}~\bibnamefont
  {Agnese}} \emph {et~al.} (\bibinfo {collaboration} {SuperCDMS}),\ }\href
  {\doibase 10.1103/PhysRevLett.112.241302} {\bibfield  {journal} {\bibinfo
  {journal} {Phys. Rev. Lett.}\ }\textbf {\bibinfo {volume} {112}},\ \bibinfo
  {pages} {241302} (\bibinfo {year} {2014})},\ \Eprint
  {http://arxiv.org/abs/1402.7137} {arXiv:1402.7137 [hep-ex]} \BibitemShut
  {NoStop}%
\bibitem [{\citenamefont {Agnes}\ \emph {et~al.}(2018)\citenamefont {Agnes}
  \emph {et~al.}}]{DarkSide:2018bpj}%
  \BibitemOpen
  \bibfield  {author} {\bibinfo {author} {\bibfnamefont {P.}~\bibnamefont
  {Agnes}} \emph {et~al.} (\bibinfo {collaboration} {DarkSide}),\ }\href
  {\doibase 10.1103/PhysRevLett.121.081307} {\bibfield  {journal} {\bibinfo
  {journal} {Phys. Rev. Lett.}\ }\textbf {\bibinfo {volume} {121}},\ \bibinfo
  {pages} {081307} (\bibinfo {year} {2018})},\ \Eprint
  {http://arxiv.org/abs/1802.06994} {arXiv:1802.06994 [astro-ph.HE]}
  \BibitemShut {NoStop}%
\end{thebibliography}%

\clearpage
\newpage
\maketitle
\onecolumngrid
\begin{center}
\textbf{\large Dark Matter Direct Detection in 3$\to$2 Process} \\
\vspace{0.1in}
{ \it \large Supplemental Material}\\
\vspace{0.05in}
{Wei Chao, \ Mingjie Jin, and \ Ying-Quan Peng}
\end{center}
\onecolumngrid
\setcounter{equation}{0}
\setcounter{figure}{0}
\setcounter{table}{0}
\setcounter{section}{0}
\setcounter{page}{1}
\makeatletter
\renewcommand{\theequation}{S\arabic{equation}}
\renewcommand{\thefigure}{S\arabic{figure}}

\subsection{Derivation of the scattering amplitude in bound state}
\label{supsec:bsf}

\begin{figure}[h]
  \centering
  \includegraphics[width=0.4\textwidth]{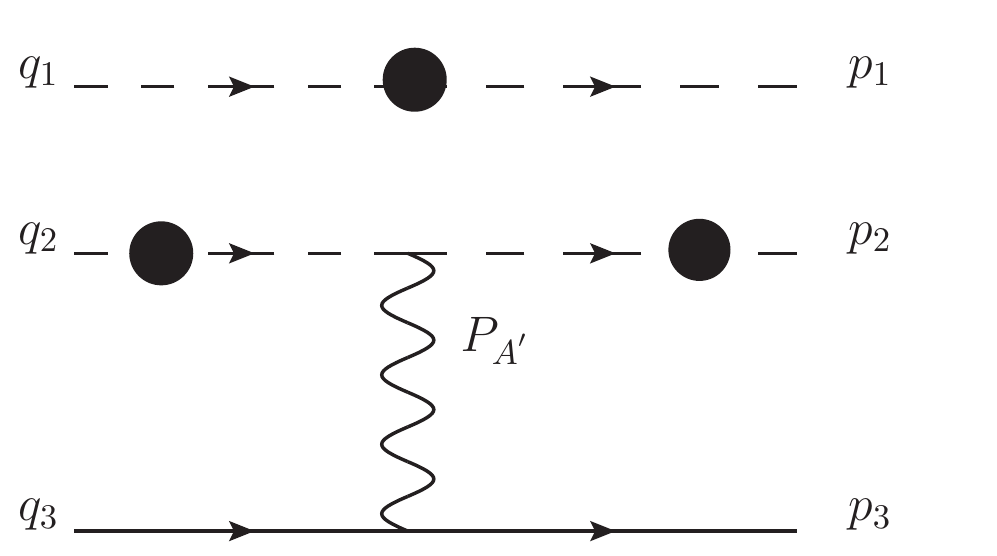}
  \includegraphics[width=0.4\textwidth]{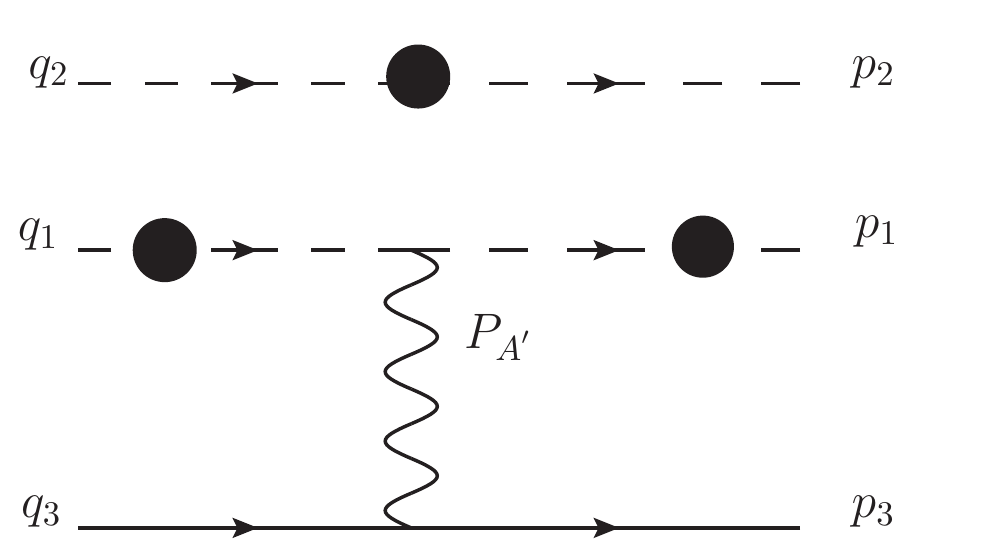}
  \caption{The leading-order contribution of the entire $\chi \chi e \to[\chi\chi]_B e$ process.}\label{BSF-leading}
\end{figure}

Following the quantum-field-theoretical procedure in Ref~\cite{Petraki:2015hla}, the leading-order contribution of the entire $\chi \chi e \to[\chi\chi]_B e$ scattering is shown in Fig.~\ref{BSF-leading}, where $q_1, q_2$ and $p_1, p_2$ are the momenta of the dark matter in the scattering states and the bound states, $q_3$, $p_3$, and $P_{A^{'}}$ are the momenta of the initial electron, the final electron and the dark photon, respectively. ``$-\bullet-$" denotes the full propagators of dark matter. 
Without loss of generality, we denote the initial dark matter as $\chi_1, \chi_2$, and their coupling coefficients with the dark photon as $g_1, g_2$ respectively. Then the contribution of Fig.~\ref{BSF-leading} is evaluated to be
{\footnotesize
\begin{align}
&(2\pi)^4 \delta^4(q_1 + q_2 + q_3 -p_1-p_2- p_3) \, i \,
{\cal C}_{{A^{'}}-\rm amp}^{(5)\mu}\tilde S_{A^{'}}(p_3-q_3)\Big[\bar{u}(p_3)(-i\epsilon e\gamma^\nu)u(q_3)\Big]
(2\pi)^4 \delta^4(q_3+P_{A^{'}}-p_3)\simeq    \nonumber
\\
&\Bigg\{-i g_1 (p_1^\mu + q_1^\mu)
\, \tilde S_1(p_1) \tilde S_1(q_1)
\, (2\pi)^4 \delta^4 (p_3-q_3 + p_1 - q_1)
\, \tilde S_2(q_2)
\, (2\pi)^4 \delta^4(p_2 - q_2) \tilde S_{A^{'}}(p_3-q_3)
(2\pi)^4 \delta^4(q_3+P_{A^{'}}-p_3)
 \nonumber
 \\
&-i g_2 (p_2^\mu + q_2^\mu)
\, \tilde S_2(p_2) \tilde S_2(q_2)
\, (2\pi)^4 \delta^4 (p_3-q_3 + p_2 - q_2)
\tilde S_1(q_1)
\, (2\pi)^4 \delta^4(p_1 - q_1)\tilde S_{A^{'}}(p_3-q_3)
(2\pi)^4 \delta^4(q_3+P_{A^{'}}-p_3)\Bigg\}
 \nonumber
 \\
&\times\Big[\bar{u}(p_3)(-i\epsilon e\gamma^\nu)u(q_3)\Big]
\: ,  \label{amplitudeA}
\end{align}}
where ${\cal C}_{{A^{'}}-\rm amp}^{(5)\mu}={\cal C}_{{A^{'}}-\rm amp}^{(5)\mu} (P_{A^{'}}, p_1,p_2,q_1,q_2) $. Integrating $P_{A^{'}}$ and $q_3$ on both sides of eq.(\ref{amplitudeA}), we have
\begin{align}
&{\cal C}_{{A^{'}}-\rm amp}^{(5)\mu}\tilde S_{A^{'}}(q_1+q_2-p_1-q_2)
\Big[\bar{u}(p_3)(-i\epsilon e\gamma^\nu)u(p_1+p_2+p_3-q_1-q_2)\Big]\simeq    \nonumber
\\
&- g_1 (p_1^\mu + q_1^\mu)
\, \tilde S_1(p_1) \tilde S_1(q_1)
\, \tilde S_2(q_2)
\, (2\pi)^4 \delta^4(p_2 - q_2) \tilde S_{A^{'}}(q_1-p_1)
\Big[\bar{u}(p_3)(-i\epsilon e\gamma^\nu)u(p_1+p_3-q_1)\Big]
 \nonumber
 \\
&- g_2 (p_2^\mu + q_2^\mu)
\, \tilde S_2(p_2) \tilde S_2(q_2)\tilde S_1(q_1)
\, (2\pi)^4 \delta^4(p_1 - q_1)\tilde S_{A^{'}}(q_2-p_2)\Big[\bar{u}(p_3)(-i\epsilon e\gamma^\nu)u(p_2+p_3-q_2)\Big]
\: . \label{amplitudeB}
\end{align}

Next we define
\begin{align}
\eta_{1,2}=\frac{m_{1,2}}{m_1+m_2}, 
\end{align}
and
\begin{align}
p_1=\eta_1P+p,\quad q_1=\eta_1K+k,\nonumber
 \\
p_2=\eta_2P-p,\quad q_2=\eta_2K-k,
\end{align}
where $P, p$ and $K, k$ are the conjugate momenta of the relevant coordinates, see chapter 3.1 in Ref\cite{Petraki:2015hla} for details. The conservation of 4-momentum gives
\begin{align}\label{4-conservation}
K+q_3=P+p_3,\quad K=P+P_{A^{'}},
\end{align}
then eq.(\ref{amplitudeB}) can be rewritten as
{\footnotesize
\begin{align}
&{\cal C}_{{A^{'}}-\rm amp}^{(5)\mu}\tilde S_{A^{'}}(P_{A^{'}})
\Big[\bar{u}(p_3)(-i\epsilon e\gamma^\nu)u(p_3-P_{A^{'}})\Big]\simeq    \nonumber
\\
&- g_1S(k;K) \Big[2\eta_1K^\mu-(\eta_1-\eta_2)P_{A^{'}}^\mu+2p^\mu\Big]
\, \tilde S_1(\eta_1P+p)
\, (2\pi)^4 \delta^4(k-p-\eta_2P_{A^{'}}) \tilde S_{A^{'}}(P_{A^{'}})
\Big[\bar{u}(p_3)(-i\epsilon e\gamma^\nu)u(p_3-P_{A^{'}})\Big]
 \nonumber
 \\
&- g_2S(k;K)\Big[2\eta_2K^\mu+(\eta_1-\eta_2)P_{A^{'}}^\mu-2p^\mu\Big]
\, \tilde S_2(\eta_2P-p)
\, (2\pi)^4 \delta^4(k-p+\eta_1P_{A^{'}})\tilde S_{A^{'}}(P_{A^{'}})\Big[\bar{u}(p_3)(-i\epsilon e\gamma^\nu)u(p_3-P_{A^{'}})\Big]
\: , \label{amplitudeC}
\end{align}}
where ${\cal C}_{{A^{'}}-\rm amp}^{(5)\mu}={\cal C}_{{A^{'}}-\rm amp}^{(5)\mu} (P_{A^{'}}, \eta_1K+k,\eta_2K-k,\eta_1P+p,\eta_2P-p) $,
$\tilde S_{A^{'}}(P_{A^{'}})=\frac{-ig_{\mu\nu}}{P^2_{A^{'}}-m^2_{A^{'}}}$ is the propagator, and we have used the definition\cite{Petraki:2015hla}
\begin{align}
S(k;K)=\tilde S_1(q_1) \tilde S_2(q_2).
\end{align}

For convenience, we define
\begin{align}
{\cal C}_{{A^{'}}-\rm amp}^{(5)\mu}\tilde S_{A^{'}}(P_{A^{'}})
\Big[\bar{u}(p_3)(-i\epsilon e\gamma^\nu)u(p_3-P_{A^{'}})\Big]=
\mathcal M^{\mu}_{\rm trans}\Big[\bar{u}(p_3)\gamma_\mu u(p_3-P_{A^{'}})\Big],
\end{align}
from eq.(\ref{amplitudeC}), we find
\begin{eqnarray}
\mathcal M^{\mu}_{\rm trans}&=&\frac{\epsilon eS(k;K)}
{P^2_{A^{'}}-m^2_{A^{'}}}
\Bigg\{g_1 \Big[2\eta_1K^\mu-(\eta_1-\eta_2)P_{A^{'}}^\mu+2p^\mu\Big]
 \tilde S_1(\eta_1P+p)
 (2\pi)^4 \delta^4(k-p-\eta_2P_{A^{'}})  \nonumber \\
&~&+g_2\Big[2\eta_2K^\mu+(\eta_1-\eta_2)P_{A^{'}}^\mu-2p^\mu\Big]
 \tilde S_2(\eta_2P-p)(2\pi)^4 \delta^4(k-p+\eta_1P_{A^{'}})\Bigg\}.
 \label{Mtrans}
\end{eqnarray}
Using the approximate result of $\mathcal M^{\mu}_{\rm trans}(\vec{p},\vec{k})$ calculated in Ref\cite{Petraki:2015hla}
\begin{eqnarray}
\mathcal M^{\mu}_{\rm trans}(\vec{p},\vec{k})=\frac{1}{ {\cal S}_0 (\vec k;K) \, {\cal S}_0 (\vec p; P)}
\int\frac{dp^0}{2\pi}\int\frac{dk^0}{2\pi}\mathcal M^{\mu}_{\rm trans},
\label{Mpq}
\end{eqnarray}
and substituting eq.(\ref{Mtrans}) into eq.(\ref{Mpq}), we obtain
\begin{eqnarray}
\mathcal M^{\mu}_{\rm trans}(\vec{p},\vec{k})&=&\frac{\epsilon eS(k;K)}
{P^2_{A^{'}}-m^2_{A^{'}}}\frac{1}{ {\cal S}_0 (\vec k;K) \, {\cal S}_0 (\vec p; P)}\times
\nonumber \\
&~&
\Bigg\{g_1 \Big[2\eta_1K^\mu-(\eta_1-\eta_2)P_{A^{'}}^\mu+2p^\mu\Big]
 \Xi_1(\vec k, \vec p; K, P) (2\pi)^3 \delta^3(\vec{k}-\vec{p}-\eta_2\vec{P}_{A^{'}}) \nonumber \\
&~&+g_2\Big[2\eta_2K^\mu+(\eta_1-\eta_2)P_{A^{'}}^\mu-2p^\mu\Big]
\Xi_2(\vec k, \vec p; K, P)(2\pi)^3 \delta^3(\vec{k}-\vec{p}+\eta_1\vec{P}_{A^{'}})\Bigg\},\nonumber
 \\ \label{Mpq2}
\end{eqnarray}
where the forms of $\Xi_1, \Xi_2$ and their non-relativistic approximations can be found in chapter 5 of Ref\cite{Petraki:2015hla}. Then the transition amplitude can be expressed in terms of the Schrodinger wavefunctions
\begin{align}
\mathcal M^i_{\vec k \to n}  &\simeq \sqrt{2\mu}
\int \frac{d^3p}{(2\pi)^3} \frac{d^3k}{(2\pi)^3}
\: \frac{\tilde \psi_n^\star (\vec p)  \: \tilde \phi_{\vec k}(\vec q)}
{\sqrt{2{\cal N}_{\vec P}(\vec p) \, 2{\cal N}_{\vec K}(\vec k)}}
\ \mathcal M^i_{\rm trans} (\vec k; \vec p) \: ,
\label{eq:Ampl BSF Inst}
\end{align}
where
\begin{align}
\frac{1}{\sqrt{2{\cal N}_{\vec P}(\vec p) \, 2{\cal N}_{\vec K}(\vec k)}} &\simeq
\frac{1}{2\mu} \left[1-\frac{\vec p^2 + \vec k^2}{4\mu^2} (1-\frac{3\mu}{M})\right] \, ,
\label{eq:Norm trans NR}
\end{align}
and
\begin{align}\label{M-mu}
\mu=\frac{m_1m_2}{m_1+m_2};\quad M=m_1+m_2,
\end{align}
are the reduced and the total masses of $\chi_1-\chi_2$, respectively.

Substituting eqs.(\ref{Mpq2})$\sim $(\ref{M-mu}) into eq.(\ref{eq:Ampl BSF Inst}), the transition amplitude is changed to be
\begin{align}
\mathcal M^i_{\vec k \to n}  &\simeq \sqrt{2\mu}
\int \frac{d^3p}{(2\pi)^3} \frac{d^3k}{(2\pi)^3}
\tilde \psi_n^\star (\vec p)  \: \tilde \phi_{\vec k}(\vec k)
\frac{1}{2\mu} \left[1-\frac{\vec p^2 + \vec k^2}{4\mu^2} (1-\frac{3\mu}{M})\right]
\frac{\epsilon e}{P^2_{A^{'}}-m^2_{A^{'}}}
\nonumber \\
&\Bigg\{g_1 \Big[2\eta_1K^i-(\eta_1-\eta_2)P_{A^{'}}^i+2p^i\Big]
 2 m_2 \left[1 + \frac{\vec p^2}{2\mu^2} (1 - \frac{2\mu}{M})\right] (2\pi)^3 \delta^3(\vec{k}-\vec{p}-\eta_2\vec{P}_{A^{'}}) +\nonumber \\
&g_2\Big[2\eta_2K^i+(\eta_1-\eta_2)P_{A^{'}}^i-2p^i\Big]
 2 m_1\left[1 + \frac{\vec p^2}{2\mu^2} (1 - \frac{2\mu}{M})\right](2\pi)^3 \delta^3(\vec{k}-\vec{p}+\eta_1\vec{P}_{A^{'}})\Bigg\},
 \label{Mkn1}
\end{align}
we only keep the next-leading-order contribution in $p$ and integrate $k$ out, then eq.(\ref{Mkn1}) is reduced to
\begin{align}
\mathcal M^i_{\vec k \to n}  &\simeq \sqrt{2\mu}\frac{\epsilon e}{P^2_{A^{'}}-m^2_{A^{'}}}
\int \frac{d^3p}{(2\pi)^3} \tilde \psi_n^\star (\vec p)  \:
\Bigg\{g_1 \Big[2\eta_1K^i-(\eta_1-\eta_2)P_{A^{'}}^i+2p^i\Big]\frac{m_2}{\mu}\tilde \phi_{\vec q}(\vec{p}+\eta_2\vec{P}_{A^{'}})
\nonumber \\
&+g_2\Big[2\eta_2K^i+(\eta_1-\eta_2)P_{A^{'}}^i-2p^i\Big]\frac{m_1}{\mu}\tilde \phi_{\vec k}(\vec{p}-\eta_1\vec{P}_{A^{'}})\Bigg\}.
 \label{Mkn2}
\end{align}
Some useful integrals are introduced in Ref\cite{Petraki:2015hla}
\begin{align}
{\cal I}_{\vec k, n} (\vec b)
&\equiv
\int \frac{d^3p}{(2\pi)^3} \:\tilde \psi_n^\star (\vec p) \: \tilde \phi_{\vec k}  (\vec p + \vec b) \: ,
\label{eq:I cal k-n def}
\\
\vec{\cal J}_{\vec k, n} (\vec b)
&\equiv
\int \frac{d^3p}{(2\pi)^3} \: \vec p \: \tilde \psi_n^\star (\vec p) \: \tilde \phi_{\vec k}  (\vec p + \vec b) \: ,
\label{eq:J cal k-n def}
\\
{\cal K}_{\vec k, n} (\vec b)
&\equiv
\int \frac{d^3p}{(2\pi)^3} \: \vec p^2 \: \tilde \psi_n^\star (\vec p) \: \tilde \phi_{\vec k}  (\vec p + \vec b) \: ,
\label{eq:K cal k-n def}
\end{align}
we can re-express eq.(\ref{Mkn1}) in terms of these integrals as follows
\begin{multline}
\mathcal M_{\vec k \to n}^j =
2\sqrt{2\mu}\frac{\epsilon e}{P^2_{A^{'}}-m^2_{A^{'}}} \ \left\{
 \frac{g_1}{\eta_1}\, {\cal J}_{\vec k, n}^j (\eta_2 \vec P_{A^{'}})
-\frac{g_2}{\eta_2}\, {\cal J}_{\vec k, n}^j (-\eta_1 \vec P_{A^{'}})
\right. \\  \left.
+ \left[g_1 (K^j - \frac{\eta_1 -\eta_2}{2\eta_1} P_{A^{'}}^j ) \: {\cal I}_{\vec k, n} (\eta_2 \vec P_{A^{'}})
+g_2 (K^j + \frac{\eta_1 -\eta_2}{2\eta_2} P_{A^{'}}^j ) \: {\cal I}_{\vec k, n} (-\eta_1 \vec P_{A^{'}})\right]
\right\} \: .
\label{eq:M SSV}
\end{multline}

Finally, according to the dark matter bound-state formation amplitude in eq.(\ref{eq:M SSV}), the total scattering amplitude squared of the $\chi \chi e \to[\chi\chi]_B e$ process can be obtained
\begin{align}\label{scat-amp1}
\overline{|\mathcal M_{\vec k {\to n}}|^2}&=\frac{1}{2}\sum_{\rm spin}\Big|\mathcal M_{\vec k \to n}^\mu\nonumber
\bar{u}(p_3)\gamma_\mu u(p_3-P_{A^{'}})\Big|^2\\
&=\mathcal M_{\vec k \to n}^\mu(\mathcal M_{\vec k \to n}^{\nu})^*
\Big[2g_{\mu\nu}(P_{A^{'}}\cdot p_3)-2P_{A^{'}\nu} p_{3\mu}-2P_{A^{'}\mu} p_{3\nu} +4p_{3\mu} p_{3\nu}\Big].
\end{align}
The Ward-Takahashi identity tells us that
\begin{align}\label{Ward-Takahashi}
(P_{A^{'}})_\mu&\mathcal M_{\vec k \to n}^{\mu}(P_{A^{'}};q_1,q_2;p_1,p_2)
\nonumber\\
&=g_1\sum^2_{i=1}\Bigg[\mathcal M_0(P_{A^{'}};q_1,q_2;p_i-P_{A^{'}})
-\mathcal M_0(P_{A^{'}};q_i+P_{A^{'}};p_1,p_2)\Bigg]
\nonumber\\
&+g_2\sum^2_{i=1}\Bigg[\mathcal M_0(P_{A^{'}};q_1,q_2;p_i-P_{A^{'}})
-\mathcal M_0(P_{A^{'}};q_i+P_{A^{'}};p_1,p_2)\Bigg]
\nonumber\\
&=\big(g_1+g_2\big)\sum^2_{i=1}\Bigg[\mathcal M_0(P_{A^{'}};q_1,q_2;p_i-P_{A^{'}})
-\mathcal M_0(P_{A^{'}};q_i+P_{A^{'}};p_1,p_2)\Bigg].
\end{align}
For the attractive DM interaction,
\begin{align}
g_1g_2<0.
\end{align}
In the case of a identical dark matter pair, we have
\begin{align}
m_1=m_2=m_\chi,
\end{align}
then $\mu$, $\eta_1, \eta_2$ and $g_1, g_2$ can be obtained
\begin{align}
&\eta_1=\eta_2=\frac{1}{2},\\
&\mu=\frac{m_\chi}{2},\\
&g_1=-g_2=g_D.
\end{align}
The Ward-Takahashi identity in eq.(\ref{Ward-Takahashi}) is reduced to
\begin{align}\label{Ward}
(P_{A^{'}})_\mu&\mathcal M_{\vec k \to n}^{\mu}(P_{A^{'}};q_1,q_2;p_1,p_2)=0,
\end{align}
which is known as Ward identity. According to the Ward identity, the 0-component of dark matter bound-state formation amplitude can be written as
\begin{align}\label{Ward}
\mathcal M_{\vec k \to n}^{0}=\frac{P_{A^{'}}^i \mathcal M_{\vec k \to n}^{i}}{P_{A^{'}}^0}.
\end{align}
Therefore, we can express the total scattering amplitude squared in the form that only contains $i, j(1, 2, 3)$-components,
{\footnotesize
\begin{align}\label{scat-ampj}
&\overline{|\mathcal M_{\vec k \to n}|^2}
\nonumber\\
&=2(P_{A^{'}}\cdot p_3)\left[\frac{|P_{A^{'}}^i \mathcal M_{\vec k \to n}^{i}|^2}{(P_{A^{'}}^0)^2}
-\mathcal M_{\vec k \to n}^j(\mathcal M_{\vec k \to n}^{j})^*\right]+
\nonumber\\
&4\left[(p^0_3)^2\frac{|P_{A^{'}}^i \mathcal M_{\vec k \to n}^{i}|^2}{(P_{A^{'}}^0)^2}
-\frac{p^0_3}{P^0_{A^{'}}}(P^j_{A^{'}}\mathcal M_{\vec k \to n}^{j})^*(p^i_3\mathcal M_{\vec k \to n}^{i})
-\frac{p^0_3}{P^0_{A^{'}}}(p^j_3\mathcal M_{\vec k \to n}^{j^*})(P^i_{A^{'}}\mathcal M_{\vec k \to n}^{i})
+(p^j_3\mathcal M_{\vec k \to n}^{j^*})(p^i_3\mathcal M_{\vec k \to n}^{i})\right].
\end{align}}

Following the calculations in Ref\cite{Petraki:2015hla}, for the capture in the ground state $\{100\}$, we keep only the leading-order terms for
${\cal I}_{\vec k, \{100\}}$ and ${\cal J}_{\vec k, \{100\}}$,
\begin{align}
\bold{\cal \vec I}_{\vec k, \{100\}} (\vec b) \simeq
\frac{2{\cal R}(\zeta)}{1+\zeta^2} \: \frac{b}{k^{5/2}}\cos\tilde{\theta},
\label{eq:I cal fin}
\end{align}
\begin{align}
\bold{\cal \vec J}_{\vec k, \{100\}} (\vec b) \simeq
\frac{{\cal R}(\zeta)}{k^{3/2}} \vec k,
\label{eq:J cal fin}
\end{align}
and the parameters $\theta$ and $\vec b$ are defined as
\begin{align}
\cos \theta =  \frac{\vec k \cdot \vec P_{A^{'}}}{|\vec k| |\vec P_{A^{'}}|}
= \frac{\vec k \cdot \vec p_3}{k |\vec p_3|} ,
\label{eq:theta}
\end{align}
with
\begin{eqnarray}
\tilde{\theta} =
\begin{cases}
\theta, \quad &\text{for } \vec b = \eta_2 \vec P_{A^{'}}
\cr
\pi + \theta, \quad &\text{for } \vec b = -\eta_1 \vec P_{A^{'}}
\end{cases}
\end{eqnarray}
other parameters can be found in Ref\cite{Petraki:2015hla}.

Under the above conditions, the dark matter bound-state formation amplitude in eq.(\ref{eq:M SSV}) is approximate to be
\begin{align}
\mathcal M_{\vec k \to n}^j =&
2\sqrt{2\mu}\frac{\epsilon e}{P^2_{A^{'}}-m^2_{A^{'}}} \ \Bigg\{
 (g_1-g_2)\frac{2{\cal R}(\zeta)}{|\vec k|^{3/2}}\, k^j \sin\theta+(g_1-g_2)
\frac{{\cal R}(\zeta)}{(1+\zeta^2)}\frac{|\vec P_{A^{'}}|}{|\vec k|^{5/2}}K^j\cos\theta
\Bigg\} \: .
\label{eq:M SSV2}
\end{align}

We calculate the total scattering amplitude in the rest frame of dark matter and electron. The relevant 4-momenta are given by
\begin{align}\label{4-momenta}
&K=\Big(2m_\chi, 0\Big),\quad k=\Big(0,\mu\vec v_{\rm rel}\Big),\quad q_3=\Big(m_e, \vec{0}\Big),
\nonumber\\
&P=\Big(E_B, -\vec{q}\Big),\quad~ p_3=\Big(E^{'}_e, \vec{q}\Big),
\end{align}
the relative velocity $\vec v_{\rm rel}$ in $k$ isn't negligible due to the non-singularity of dimensionless parameter $\zeta=\frac{\alpha}{|\vec v_{\rm rel}|}$. Then the amplitude in eq.(\ref{eq:M SSV2}) is reduced to
\begin{align}
\mathcal M_{\vec k \to \{100\}}^j =&
2\sqrt{2\mu}\frac{\epsilon e}{P^2_{A^{'}}-m^2_{A^{'}}} \ \Bigg\{
 (2g_D)\frac{2{\cal R}(\zeta)}{|\vec k|^{3/2}}\, k^j \sin\theta\Bigg\} \: .
\label{M-final}
\end{align}

\subsection{The matrix element squared for DM-electron scattering}
\label{supsec:msq_dme}

In this section we calculate the analytical expressions of matrix elements squared of DM-electron scattering for dark photon and bound state final state. According to feynman diagrams in Fig.~\ref{fig:3to2}, the matrix element squared for scalar DM is,

\bea
\overline{|\mathcal{M}|^2}_S&=&\frac{64 \pi \alpha g_D^4 m_e \epsilon^2 [2 m_\chi E_{A'} (m_\chi+m_e)-m_{A'}^2 (E_{A'}-m_\chi+m_e)]}{m_{A'}^2 (-2 m_e E_{A'}+m_{A'}^2+4 m_\chi m_e)^2}\label{eq:ms_scalar}
\label{eq:ms_photon}
\eea
where the total energy of dark photon $E_{A'}=\sqrt{m_{A'}^2+q^2}\simeq q$. For the bound state DM, combing Eq.~\ref{scat-ampj}, Eq.~\ref{4-momenta} and Eq.~\ref{M-final}, we obtain
\begin{align}
\overline{|\mathcal M_{\vec k \to \{100\}}|^2}
=&\frac{\epsilon^2 e^2(2g_D)^2}{\left[-4m_em_\chi+2m_e\sqrt{\vec q^2+M_{B\{100\}}^2}-m^2_{A^{'}}\right]^2}
\frac{64}{|\vec v_{\rm rel}|}|{\cal R}(\zeta)|^2\times
\nonumber\\
&\Bigg\{\left((2m_\chi-E_{B\{100\}})E^{'}_e-{\vec q}^2\right)
\left[\frac{\vec{q}^2\cos^2\theta\sin^2\theta}{(2m_\chi-E_{B\{100\}})^2}
-\sin^2\theta\right]
\nonumber\\
&+2(E^{'}_e)^2\left[\frac{\vec{q}^2\cos^2\theta\sin^2\theta}{(2m_\chi-E_{B\{100\}})^2}\right]
-\frac{4E^{'}_e}{2m_\chi-E_{B\{100\}}}\vec{q}^2\cos^2\theta\sin^2\theta
+2\vec{q}^2\cos^2\theta\sin^2\theta\Bigg\}.
\label{eq:ms_bdm}
\end{align}

\subsection{Differential ionization rate for DM-electron scattering}
\label{supsec:dm_e}
The cross section for $3\to2$ process is written as follows,
\begin{align}
\langle \sigma v^2 \rangle &= \frac{1}{4 E_{A^{'}} E_e'} \int \frac{d^3 q}{(2\pi)^3} \frac{d^3 k'}{(2\pi)^3}
 \frac{1}{8 E^2_\chi E_e} (2 \pi)^4 \delta(E_i - E_f) \delta^3(\vec k + \vec q - \vec k')
\overline{|\mathcal M(q\,)|^2} \times | f(\vec q \,)|^2  ,\\
&=\frac{1}{32 E_{A^{'}} E_e' E^2_\chi E_e} \int \frac{d^3 q}{(2\pi)^3} 2\pi \delta(\Delta E - 2 m_\chi + \sqrt{q^2 + m^2_{A^{'}}})
\overline{|\mathcal M(q\,)|^2}  | f(\vec q \,)|^2
\label{eq:free-cross-section}
\end{align}

Following the procedure in Ref.~\cite{Dror:2020czw}, the event rate is derived by,
\begin{align}
R &= \left(\frac{\rho_\chi}{m_\chi}\right)^2 \int d^3 v \, g_\chi(v) \, \langle \sigma v^2 \rangle \, \\
&=\frac{\rho_\chi^2}{32 m^2_\chi E_{A^{'}} E_e' E_e E^2_\chi } \int \frac{d^3 q}{(2\pi)^2} d^3 v \, g_\chi(v) \, \delta(\Delta E - 2 m_\chi + \sqrt{q^2 + m^2_{A^{'}}})
\overline{|\mathcal M(q\,)|^2}  | f(\vec q \,)|^2 , \\
&=\frac{\rho_\chi^2}{32 m^2_\chi E_{A^{'}} E_e' E_e E^2_\chi } \int \frac{d^3 q}{(2\pi)^2} \delta(\Delta E - 2 m_\chi + \sqrt{q^2 + m^2_{A^{'}}})
\overline{|\mathcal M(q\,)|^2}  | f(\vec q \,)|^2 ,
\label{eq:app_eventrate}
\end{align}
where $g_\chi(\vec v)$ is the distribution function of dark matter velocity. After replacing the form factor $f(q) \to f_{\rm ion}(k', q)$ in Eq. (\ref{eq:app_eventrate}) by applying the Eq. (A.21) from Ref.~\cite{Dror:2020czw}, we integrate $q$ and then the differential ionization rate is written as
\begin{align}\label{R2}
\frac{d R_{3\to2}}{d E_R}
=\sum_{n, l}\frac{N_{\rm T} \rho^2_\chi}{128 \pi m^4_\chi m_e E'_e E_R}q
\overline{|\mathcal M(q\,)|^2} | f^{n,l}_{\rm ion}(k', q )|^2
\end{align}
where we take the approximation $E_\chi\approx m_\chi, E_e \approx m_e\footnote{The initial electron should be described by $E_e=m_e-E_B$ where $E_B$ is the binding energy of atomic electron. Since the $E_B \ll m_e$, thus we take the approximation $E_e \approx m_e$ for convenience.}.  E_e'=m_e+2 m_\chi - \sqrt{q^2 + m^2_{A'}}$, $E_{A'} = \sqrt{q^2 + m^2_{A'}}\simeq q$, $k'=\sqrt{ 2 m_e E_R}$, where $E_R$ is the recoil energy of electron. The Eq.~\ref{R2} is also applicable to the final state being bound state DM.

\end{document}